\documentclass[a4paper]{article}

\usepackage{graphicx}
\usepackage{amsmath}
\usepackage{amsfonts}

\newcommand\N{\mathbb{N}}
\newcommand\Z{\mathbb{Z}}
\newcommand\R{\mathbb{R}}
\newcommand\C{\mathbb{C}}
\newcommand\Q{\mathbb{Q}}
\newcommand\T{\mathbb{T}}

\def\A{{\tt A}}
\def\s{{\tt S}}
\def\I{{\tt I}}
\def\O{{\rm O}}
\font\ninerm=cmr9

\begin{document}

\centerline {\bf Darboux inversions of the Kepler problem}

\bigskip
\centerline {by Alain Albouy$^1$ and Lei Zhao$^2$}

\medskip
\centerline {$^1$ IMCCE, UMR 8028,}
\centerline {77, avenue Denfert-Rochereau}
\centerline {F-75014 Paris}
\centerline {Alain.Albouy@obspm.fr}

\medskip
\centerline {$^2$ Institute of Mathematics}
\centerline {University of Augsburg}
\centerline {Augsburg, Germany}
\centerline{Lei.Zhao@math.un-augsburg.de}

\bigskip

{\ninerm This work is dedicated to the memory of Alexey V.\ Borisov. By his impressive work he recalled us that the interesting discoveries are often old. They deserve to be studied, shared and continued, even if our understanding is delayed by our distance to the great minds of the past.}

\bigskip

{\bf Abstract.} While extending a famous problem asked and solved by Bertrand in 1873, Darboux found in 1877 a family of abstract surfaces of revolution, each endowed with a force function, with the striking property that all the orbits are periodic on open sets of the phase space. We give a description of this family which explains why they have this property: they are the Darboux inverses of the Kepler problem on constant curvature surfaces. What we call the Darboux inverse was briefly introduced by Darboux in 1889 as an alternative approach to the conformal maps that Goursat had just described.

\bigskip

{\bf {}1. Introduction}

In the Kepler problem, all the choices of an initial position together with an initial velocity lead to a periodic orbit, provided that the energy is negative and the angular momentum is non-zero.

Bertrand showed in 1873 that among the force functions in the plane which are function of the distance $r$, only the Kepler problem, with force function $m/r$, where $m>0$, and the Hooke problem, with force function $-mr^2$, have such an astonishing property.

Darboux extended Bertrand's question in 1877, by assuming that the configuration space is a surface of revolution instead of a plane. He found that if the surface and the force function are real analytic, if the force function is non-constant, then all the systems answering the question belong to two families, a 2-parameter family and a 3-parameter family (see Theorem {}7.5). Indeed, under a rationality condition, a Lagrangian system still solves the question if we replace the surface of revolution by another one which is locally isometric. Darboux described this possibility with a further rational parameter $\mu$.

Darboux identified the first family as the generalizations to surfaces of constant curvature, due to P.\ Serret in 1859, of the Kepler problem. Indeed, only the positive curvature case was described at that time. Lobachevsky had introduced in 1834 (see [Che]) a law of gravitation in the negative constant curvature space. But apparently nobody had solved the corresponding Kepler problem before Killing in 1885. However, this problem is easily found in Darboux's first family (see [ZKF], see {}8.2).

The generic surfaces of Darboux's second family remained unidentified as far as we know, even if [ZKF] proved that their curvature is varying in a monotone way along the surface (see Theorem {}8.1). We prove that they are the Jacobi-Maupertuis surfaces of the Lagrangian systems of the first family. We also need to tell which is the force function on each surface. We may describe at once the surface and the force function as follows (see our final Theorem {}9.3): {\it The second family consists in the Darboux inverses of the Lagrangians of the first family.}

We have to explain what is this forgotten transformation which we call the Darboux inversion of a quadratic Lagrangian. If 
$T$ is a quadratic homogeneous ``kinetic energy'', $U$ a force function, the first Darboux inverse of the quadratic Lagrangian $T+U$ is the quadratic Lagrangian $UT+1/U$. The second Darboux inverse is $-UT+1/U$. Their fundamental property is the following. The Lagrange equations of $T+U$ and of $UT+1/U$ have, up to reparametrization, the same solutions of positive energy (see Theorem {}6.8). The Lagrange equations of $T+U$ and of $-UT+1/U$ have, up to reparametrization, the same solutions of negative energy. In both cases the solutions should indeed be stopped at $U=0$. Note that if we change $U$ into $U+h$, where $h$ is a constant, we get different Darboux inverses. This free parameter $h$ explains why Darboux's second family has one parameter more than his first family.

Darboux discovered this inversion in 1889 while reacting to a note by Goursat which he had just presented to the French Acad\'emie des Sciences. Goursat was rediscovering the duality of central forces (see Proposition {}2.8), first published by MacLaurin in 1742.
The most famous example is the map $\C\to\C$, $z\mapsto z^2$, which sends the orbits of the attractive Hooke problem onto the orbits of the Kepler problem with negative energy (see Figure 5). Darboux's inversion gives the same result in three steps: starting with the Kepler problem in the plane, Darboux changes the metric tensor by applying a conformal factor, the Jacobi-Maupertuis factor $U=1/r$.
The plane becomes an ordinary cone with angle $\pi/3$ at the apex, which is called the Kepler cone.
Darboux endows this cone with the force function $-1/U$, and thus obtains the Lagrangian $UT-1/U$. Finally, an isometric double covering regularizes the apex and gives the attractive Hooke problem. Other isometric coverings also regularize the geometry and the dynamics in many similar examples (see {}5.1, Figure 7 and Proposition {}6.9). The Kepler problem in the 3-dimensional Euclidean space also possesses Darboux inverses, but the subsequent double covering is topologically impossible (compare [Le5], [KSt], see our timeline in \S {}10).

Note that the Lagrangians $UT-1/U$ and $-UT+1/U$ have the same dynamics. In order to reduce the number of cases in the discussion of signs, and to have a definition which keeps the same simplicity when passing to the indefinite Lagrangians, we propose the formula $-UT+1/U$ for the second Darboux inverse (see Definition {}6.7).

It is easy to write a family of Lagrangians which all lead to the same Jacobi-Maupertuis metric tensor (see Remark {}4.5). They will consequently have the same solutions, up to reparametrization, {\it on a given energy level}. The Darboux inversion has a stronger property: on all the energy levels with energy of a given sign, the solutions are, up to reparametrization, the same. A Darboux inversion will consequently preserve the periodicity of the solutions in an open set of the phase space. It will send a Lagrangian with an open set filled up with periodic orbits on another Lagrangian with the same property.

In a previous work [AlZ], we have defined the Lambert-Hamilton property, which describes systems similar to the Kepler problem, in the sense that they have a property similar to the classical Lambert theorem. If a Lagrangian satisfies the Lambert-Hamilton property, its Darboux inverse should also satisfy it, since the arcs going from a point A to a point B are just reparametrized, and since the Maupertuis action is essentially the same according to Darboux's proof of Theorem {}6.1. We have proved in [AlZ] that the systems of the first Darboux family satisfy the Lambert-Hamilton property. We should consequently expect that the systems of the second family also satisfy the property. We have proved it in [AlZ] in the constant curvature case. The other cases should yet be examined carefully due to the conditions $U\neq 0$ and $h\neq 0$ which appear in the Darboux inversion, and which may restrict the family of arcs from a point A to a point B.

In the same previous work [AlZ], we have shown the invariance of the Lambert-Hamilton property under Appell's central projection. This classical projection was introduced in 1890 by Appell, a year after the Darboux inversion, and was also motivated by considerations about the Kepler problem and the Hooke problem. Appell's projection may also transform a natural Lagrangian into a natural Lagrangian with the same solutions up to reparametrization (see [BoM]). The nearly simultaneous discovery of Appell's and Darboux's transformations stimulated the theory of the ``transformation of the equations of dynamics'' (see e.g.\ [Tho], see our timeline in \S {}10).

Our present result may be summarized as follows: the central Lagrangians (Definition {}7.1) having Darboux's closed orbit property are the images of the Kepler problem by these ``transformations''. About the Lambert theorem, which is a quite different topic, it may be asked if the central Lagrangians having the Lambert-Hamilton property are also the images of the Kepler problem by these ``transformations''.

\bigskip

{\bf {}2. The duality of homogeneous force functions}

Consider, in the Euclidean plane $\R^2=\O xy$, a homogeneous force function $U:\R^2\setminus\{\O\}\to \R$, $q\mapsto m r^{2k}$. Here, $q=(x,y)$ is the position vector, $r^2=x^2+y^2$, $k\in\R$ and $m\in\R$. The force function $U$ extends to the origin ${\O}$ as a real analytic function in the plane if and only if $k\in\N$. We will give a consequence of this extension at the end of this section. To $U$ is associated the differential
system $\ddot q=\nabla U|_q$, where $\nabla U|_q=2km r^{2k-2}q$ is the gradient of $U$ at $q$, i.e., the central force field, $\ddot q=d^2q/dt^2$ is the acceleration vector, and $t\in \R$ is the time. A typical solution $t\mapsto q(t)$ draws a quasi-periodic ``rosette'', with a constant angle $\Theta$ between two successive pericenters.

In this section we look for {\it simple transformations sending a solution of such a system onto a solution of another such system.} We will assume that the transformation multiplies the polar angle $\theta$ by a number, thus respecting the constancy of the angle between the pericenters of a rosette. The transformation may be multivalued. Also, {\it we allow a change of the time parameter.} Let us mention a transformation sending solutions onto reparametrized solutions of the same system: the rescaling $(q,\dot q)\mapsto(\lambda q, \lambda^{k}\dot q)$, where $\lambda>0$.

\bigskip
\centerline{\includegraphics[width=40mm]{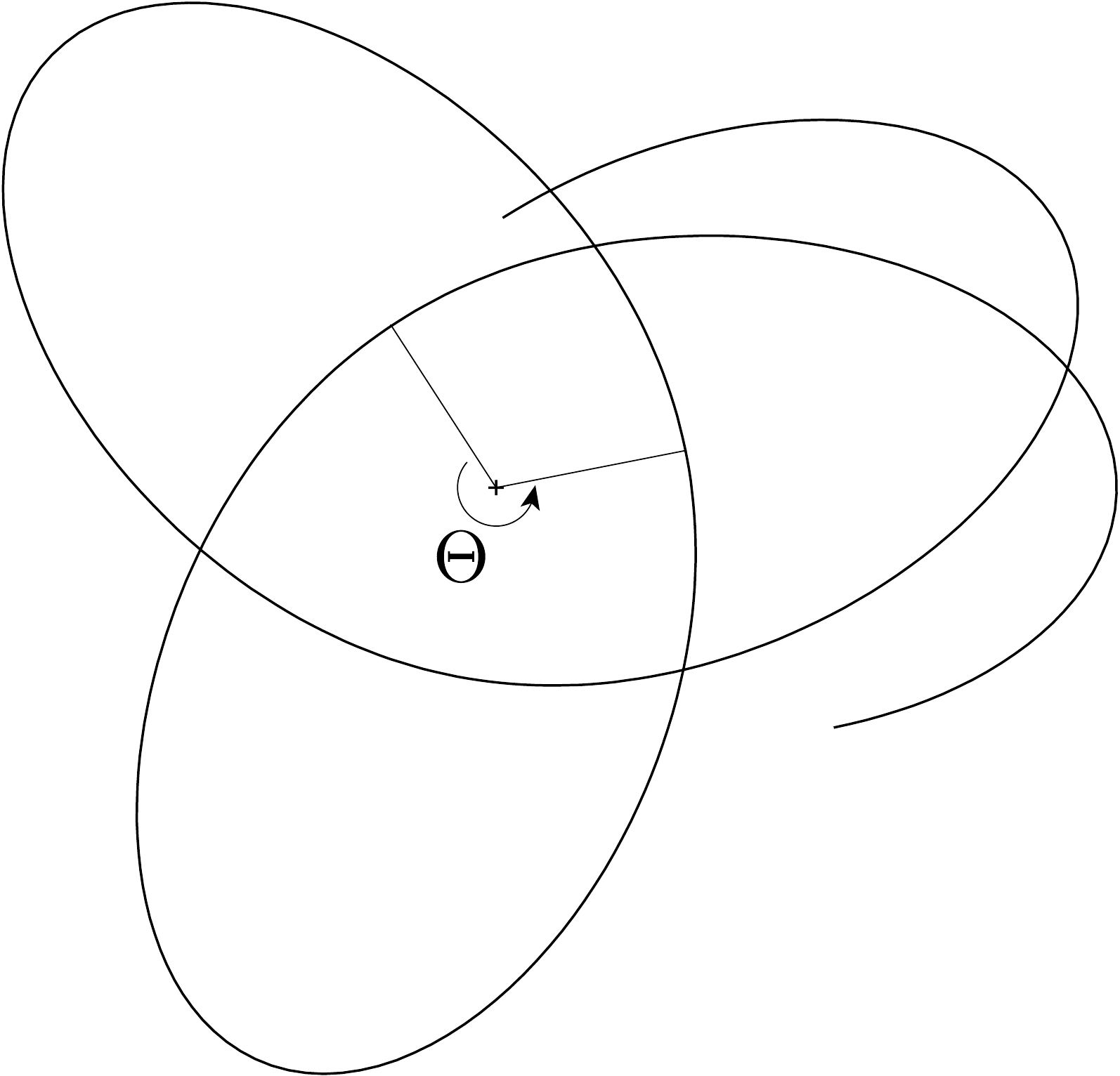}}

\nobreak
\centerline{Figure 1. A quasi-periodic rosette on an interval of time}
\bigskip

{\bf The Clairaut-Binet variables.} Let $\R^+=(0,+\infty)$. Consider the more general force function $U=-F(r^2)/2$, where $F:\R^+\to \R$ is a smooth function, associated to the same equation $$\ddot q=\nabla U|_q.\eqno({}2.1)$$ Let $f=F'$ be the derivative of $F$. Then $\ddot q=-f(r^2)q$. The angular momentum $C=x\dot y-y\dot x$ and the energy $h=\|\dot q\|^2/2-U$ are constants of motion. The {\it Clairaut-Binet variables} are $u=1/r$ and the polar angle $\theta$.
Since $C=r^2\dot\theta$ and $\dot r=-Cu'$, where $u'=du/d\theta$, we have the classical expressions:
$$\|\dot q\|^2=\dot r^2+(r\dot\theta)^2=\dot r^2+\frac{C^2}{r^2}=C^2(u'^2+u^2).\eqno({}2.2)$$
Differentiating the energy relation $$h-\frac{1}{2}F\Bigl(\frac{1}{u^{2}}\Bigr)=\frac{C^2}{2}(u'^2+u^2)\eqno({}2.3)$$
with respect to $\theta$ and simplifying $u'$ we get the familiar equation of motion in the Clairaut-Binet variables
$$\frac{f}{u^3}=C^2(u''+u).\eqno({}2.4)$$
We observe that a solution $\theta\mapsto u$ of $({}2.3)$ gives a solution $t\mapsto q$ of $({}2.1)$ with angular momentum $C$ and energy $h$. The time parameter $t$ is defined, up to an additive constant, by the relation $\dot\theta=Cu^2$.

 We now restrict to the homogeneous force functions by setting $F(s)=-2m s^k$, and consequently $f(s)=-2km s^{k-1}$. We rewrite $({}2.3)$ as
$$h+m u^{-2k}=\frac{C^2}{2}(u'^2+u^2).\eqno({}2.5)$$
If a triplet $(h,m,C^2)$ defines a non-empty curve $({}2.5)$ in the variables $(u,u')\in \R^+\times\R$, then it specifies a solution of the central force system, up to translations of the time $t$ and of the polar angle $\theta$. Any multiple $(\lambda h,\lambda m,\lambda C^2)$, with $\lambda>0$, defines the same curve, but the corresponding solutions of $({}2.1)$ are time-reparametrized, due to the new angular momentum $\sqrt{\lambda}C$ which gives a new relation between the polar angle and the time.
Let us also mention that the triplet $(\lambda^{-2k}h,m,\lambda^{-2-2k} C^2)$ specifies the rescaled solution $\lambda u$.

\bigskip

{\bf How to map a solution onto a solution.} Our question will be: when is a solution of such a central force system mapped onto a time-reparametrized solution of another such central force system by a ``homogeneous mapping'' $(\theta,u)\mapsto (\omega,v)$ where $\theta=\alpha\omega$ and $u=v^\beta$? Here $\alpha\neq 0$ and $\beta\neq 0$ are two unknown real parameters. The image of $({}2.5)$ through such a mapping satisfies: 
$$h+m v^{-2\beta k}=\frac{C^2}{2}\Bigl(\bigl(\frac{\beta v^{\beta-1}v'}{\alpha}\bigr)^2+v^{2\beta}\Bigr),$$
where $'$ now stands for the differentiation with respect to $\omega$. Multiplying this equation by $v^{2-2\beta}$ we get
$$hv^{2-2\beta}+m v^{2-2\beta(k+1)}=\frac{C^2}{2}\Bigl(\bigl(\frac{\beta v'}{\alpha}\bigr)^2+v^2\Bigr).\eqno({}2.6)$$
We look for parameters $(h,m,C^2,\alpha,\beta,k)$ such that $({}2.6)$ has the same form as $({}2.5)$. We assume that $k\neq 0$ and $k\neq -1$, so that the three monomials in $v$ all have distinct powers. To get the same form as $({}2.5)$, the $v^2$ term should have the same coefficient as the $v'^2$ term, which gives $\alpha=\pm\beta$. We can choose $\alpha=\beta$ since the change of sign of $\alpha$ defines a trivial reflection. 
After the standard identification $\O xy=\C$, the mapping is now $z\mapsto z^{1/\beta}$,
where $z=x+iy=q$ is the position. In the same complex notation, equation $({}2.1)$ becomes
$$\ddot z=2km |z|^{2k-2}z.\eqno({}2.7)$$
In particular, we see that {\it the mapping is conformal}, since it is a complex analytic mapping.
Now, in $({}2.5)$, one of the three monomials is constant. Consequently, in $({}2.6)$, one of the monomials should be constant. If $\beta\neq 1$, then it can only be the second one. This gives
$$\beta=\frac{1}{k+1},\qquad l=\beta-1,\qquad l+1=\frac{1}{k+1},\qquad hv^{-2l}+m =\frac{C^2}{2}(v'^2+v^2).\eqno({}2.8)$$
Clearly, we could also multiply the triplet of parameters $(h,m,C^2)$ by a same positive number, which, as we already observed, would only change the time-parametrization of the solutions. But let us begin with this mere exchange of parameters $(h,m,C^2)\mapsto (m,h,C^2)$, which changes $({}2.5)$ into $({}2.8)$ when we also change the exponent $k$ into $l$.

{\bf {}2.1. Lemma.} Let $\I\subset\R$ be an open interval and $\C^*=\C\setminus\{0\}$. A solution $\I\to \C^*$, $t\mapsto z$ of the system with force function $m r^{2k}$, having energy $h$ and angular momentum $C$, where $(m,h,C,k)\in\R^4$, $k\neq -1$, is mapped onto a reparametrized solution in $\C^*$ of the system with force function $h r^{2l}$, where $(k+1)(l+1)=1$, having energy $m$ and angular momentum $\pm C$, through the conformal multivalued map $\C^*\to\C^*$, $z\mapsto z^{k+1}$.

{\bf Proof.} If $C\neq 0$, it is enough to deduce $({}2.8)$ from $({}2.5)$, as we just did. If
$C=0$ then the solution is on a ray from $\O$. It is easy to check the statement in the trivial cases $m=0$ and $k=1$. In the other cases, the solution possesses at most one stationary point, i.e., a position with zero velocity. At such position the velocity changes sign, due to non-zero acceleration. We claim the correspondence up to reparametrization. So, it is enough to check that, if it exists, the stationary point is mapped onto a stationary point, and that the solution is mapped onto the correct side of the stationary point. According to $({}2.2)$, the right-hand side of $({}2.5)$ is the kinetic energy. It is $\dot r^2/2$ when $C=0$. The multiplication by $v^{2-2\beta}$ which gives $({}2.6)$ also shows that a zero $u_0$ of $h+m u^{-2k}$ is mapped onto a zero $v_0$ of $hv^{-2l}+m$, and that the intervals where these quantities are positive correspond to each other. QED

{\bf {}2.2. Remark.} We write $\pm C$ in the statements. If $k+1<0$, a solution with $C>0$ is mapped onto a solution with $C<0$. If we prefer $C>0$, we may choose to reverse the motion on the image of a solution.

{\bf {}2.3. Definition.} When a parametrized path $t\mapsto q(t)$ is {\it reparametrized}, i.e., when a second parametrized path is defined as $\tau\mapsto q\bigl(t(\tau)\bigr)$, we call the {\it ratio of velocities} at a point $q$ the quotient of the second velocity vector $dq/d\tau$ by the first velocity vector $dq/dt$.

{\bf {}2.4. Lemma.} Consider the two solutions compared in Lemma {}2.1. The ratio of velocities is $|z|^{-2k}/|k+1|$.

{\bf Proof.} Let $w=z^{k+1}$. We consider the three velocity vectors $\dot z=dz/dt$, $\dot w=(k+1)z^k\dot z$ and $dw/d\tau$ where $\tau$ is the time parameter of the motion under the force function $hr^{2l}$. As we multiplied the kinetic energy by $v^{2-2\beta}=u^{2k}$ to get $({}2.6)$, we have $|dw/d\tau|=u^k|\dot z|$. But $|k+1||\dot z|=u^k|\dot w|$. So $dw/d\tau=u^{2k}\dot w/|k+1|$. QED

{\bf {}2.5. Remark.} The correspondence of the Lemmas is encoded by the exchange of parameters $(h,m,C^2)\mapsto (m,h,C^2)$.
 The ratio of velocities is simple: it depends on the position only, and not even on the parameter $m$. But at the image, the force function $hr^{2l}$ depends on the energy $h$ of the chosen solution. {\it We may instead map all the solutions of positive energy of a system onto solutions of only one other system.} It is enough to consider the correspondence encoded by $(h,m,C^2)\mapsto (m/h,1,C^2/h)$. This gives the following proposition.

{\bf {}2.6. Proposition (MacLaurin).} All the solutions in $\C^*$, of positive energy, of the system with force function $m r^{2k}$, with $(m,k)\in\R^2$, $k\neq -1$, are mapped through the conformal multivalued map $z\mapsto z^{k+1}$ onto reparametrized solutions of the system with force function $r^{2l}$, where $(k+1)(l+1)=1$. A solution of energy $h>0$ and angular momentum $C$ is mapped onto a solution of energy $m/h$ and angular momentum $\pm C/\sqrt{h}$. The ratio of velocities is $h^{-1/2}|z|^{-2k}/|k+1|$.

As the energy is the sum of the kinetic energy and the potential energy $-m r^{2k}$, all the solutions in $\C^*$ have positive energy if $m<0$. There are solutions of negative energy in the case $m>0$. They may be mapped onto solutions of only one fixed system. We use the correspondence $(h,m,C^2)\mapsto (-m/h,-1,-C^2/h)$.

{\bf {}2.7. Proposition (MacLaurin).} All the solutions of negative energy of the system with force function $m r^{2k}$, $m>0$, $k\neq -1$, are mapped through the conformal multivalued map $z\mapsto z^{k+1}$ onto reparametrized solutions of the system with force function $-r^{2l}$, where $(k+1)(l+1)=1$. A solution of energy $h<0$ and angular momentum $C$ is mapped onto a solution of energy $-m/h>0$ and angular momentum $\pm C/\sqrt{-h}$. The ratio of velocities is $(-h)^{-1/2}|z|^{-2k}/|k+1|$.

{\bf {}2.8. Proposition (MacLaurin).} All the solutions of zero energy of the system with force function $m r^{2k}$, $m>0$, $k\neq -1$, are mapped through the conformal multivalued map $z\mapsto z^{k+1}$ onto straight lines.

The previous propositions and lemmas define a {\it duality} between homogeneous central force functions. The force functions proportional to $r^{2k}$ are dual to the force functions proportional to $r^{2l}$, where $1/k+1/l=-1$, and the correspondence is the map $z\mapsto z^{k+1}$ or its inverse $z\mapsto z^{l+1}$.
One of the most striking consequence of this duality is that it sometimes gives a {\it regularization} of system $({}2.7)$ at the singularity $z=0$, called the collision. If $k\in\N$, the force function $r^{2k}$ is smooth at the origin. The solutions which we considered on $\C^*$ are indeed well-defined on $\C$. Some regularity should be expected for the dual exponent. The example $k=1$ is well-known to give a regularization of the Kepler problem. From a duality result similar to the above three Propositions, McGehee deduced the following result. We invite the reader to consult [McG] for a definition of {\it block regularizable}.

{\bf {}2.9. Theorem (McGehee).} The singularity set for the planar system with force function $r^{2k}$, $k<0$, restricted to any given energy level set, is block regularizable if and only if $k=1/(l+1)-1$, where $l\in\N$.

\bigskip

{\bf {}3. Main examples and early approach}

MacLaurin discovered in 1742 the duality of homogeneous central force laws explained in the previous section. In 1720, he had published his {\it Geometria organica}, which contains in the second part a Proposition XXII which is essentially equivalent to our Proposition {}2.8, and that we restate as follows.

{\bf {}3.1. Proposition (MacLaurin).} If a body moves under a central force along the planar curve with polar equation $r^n=\cos n\theta$, $n\in\R^*$, then this force is, along the curve, proportional to $r^{-2n-3}$.

\bigskip

\bigskip
\centerline{\includegraphics[width=40mm]{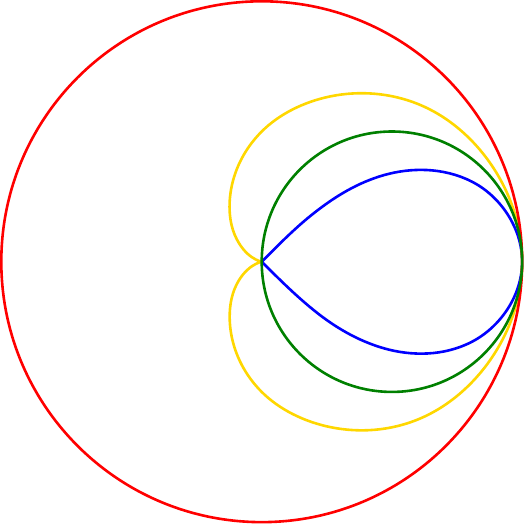}}

\nobreak
\centerline{Figure 2. Solution of zero energy starting with zero radial velocity}

\nobreak
\centerline{Lemniscate: $U=r^{-6}$, circle: $U=r^{-4}$, cardioid: $U=r^{-3}$, circle: $U=r^{-2}$}
\bigskip

In other words, setting $n=-k-1$, the differential equation of motion should be $({}2.1)$ with $U=m r^{2k}$ and $q=(r\cos\theta,r\sin\theta)$.
In the complex variable $z=re^{i\theta}$, the equation of the curve is $\Re(z^{k+1})=1$, which agrees with Proposition {}2.8. The solution on such a curve has zero energy. Consequently, $m>0$. The curves with equation $r^n=\cos n\theta$ are called the {\it sinusoidal spirals}. They were introduced by MacLaurin, also in the second part of his {\it Geometria organica}, in Proposition XIV. The curve is a parabola with focus at the origin if $n=k=-1/2$, since squaring $r^{-1/2}=\cos (\theta/2)$ gives the familiar polar equation $1/r=(1+\cos\theta)/2$. Thus Proposition {}3.1 generalizes the parabolic motion under Newton's law of forces. The case $n=1$, $k=-2$ is the circular motion toward a center located on the circle, also due to Newton in his {\it Principia}, Proposition 7, Corollary 1. MacLaurin presented other examples as corollaries to his Proposition XXII. The subsequent propositions consider the motion in a resisting medium, another subject of the {\it Principia}. These mechanical contents of {\it Geometria organica} are not well-known. We learned about them in [Coo]. They are described in their context in [Gui], \S 2.3.

\bigskip

\bigskip
\centerline{\includegraphics[width=110mm]{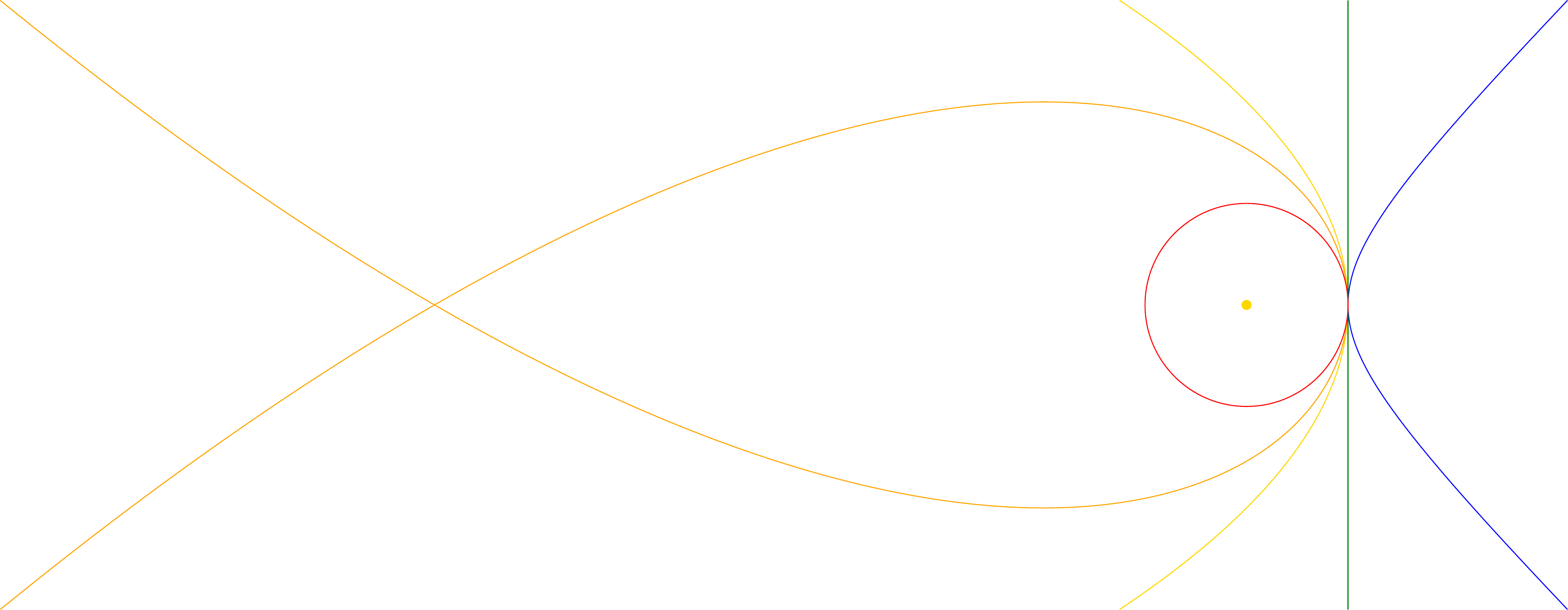}}

\nobreak
\centerline{Figure 3. Continuation. Circle: $U=r^{-2}$, Tschirnhausen cubic: $U=r^{-4/3}$,}

\nobreak
\centerline{parabola: $U=r^{-1}$, line: $U=1$, equilateral hyperbola: $U=r^2$}
\bigskip

In 1742, MacLaurin came back to the central forces in his {\it Treatise of fluxions}. At \S 451 and again in \S 875, he states the duality in terms of polar coordinates. He does not use the power of complex numbers, but multiplies the angle and takes a power of the distance accordingly. The following remarkable proof is given after the second statement. Here A is a point which belongs to both solutions, M is the moving point of the first solution, L of the second solution, S is the center. MacLaurin writes ${\rm SA}=1$, ${\rm SM}=x$, ${\rm SL}=r$, $r=x^{(m+3)/2}$, ${\rm ASL}:{\rm ASM}::m+3:2$, which means that the ratio of the angles is $(m+3)/2$.

 \bigskip
\centerline{\includegraphics[width=50mm]{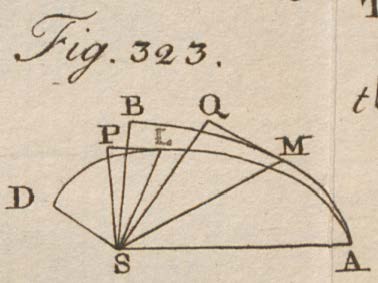}}

\nobreak
\centerline{Figure 4. The curve AMB and its complex power, the curve ALD}

\nobreak
\centerline{Figure of [Ma2], source: ETH-Bibliothek}
\bigskip

``For let SQ and SP be perpendicular to the respective tangents of AM and AL in Q and P, ${\rm SQ}=y$, and ${\rm SP}=p$.
Then, by the supposition $\dot y/(y^3\dot x)=ex^m$, where $e$ represents an invariable quantity. By finding the fluents $1/y^2=2K-2ex^{m+1}/ (m+1)$, where $K$ denotes an invariable quantity according to art.\ 735. The triangles SMQ, SLP being similar (art.\ 394) it follows that $1/p^2=x^2/(r^2y^2)=$ (because $r^2=x^{m+3}$) $1/(y^2x^{m+1})=2K/x^{m+1}-2e/(m+1)=2Kr^{-(2m+2)/(m+3)}-2e/(m+1)$, and $\dot p/p^3\dot r=(4m+4)Kr^{(-3m-5)/(m+3)}/(m+3)$, or as the power of $r$ of the exponent $4/(m+3)-3$.''

Let us explain and compare. Our map is $z\mapsto z^{k+1}$ so MacLaurin exponent $(m+3)/2=k+1$. The quantity ${\rm SQ}=y$
is, on a solution, as explained by Newton, inversely proportional to the velocity $\|\dot q\|$, since by orthogonality the constant angular momentum is $C=y\|\dot q\|$. The first equation $\dot y/y^3=ex^m\dot x$ expresses that the derivative of the kinetic energy is proportional to the derivative of the potential energy, and the second equation is the conservation of the energy.
The similarity of the triangles is needed in order to give the relation between $y$ and $p$. The proof in \S 394 is long. But, as the complex map is conformal, we have a shorter proof: the angles SMQ and SLP are the same. Since the angles SQM and SPL are also the same, being $\pi/2$, we have the similarity. The proof by MacLaurin is consequently just a few lines long.

 \bigskip
\centerline{\includegraphics[height=30mm]{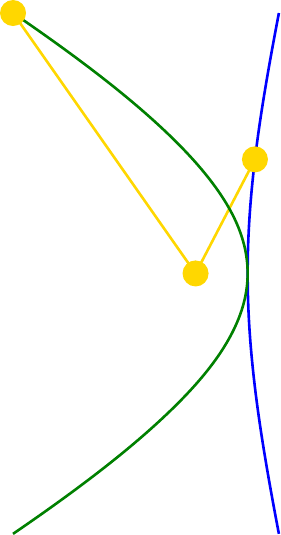}$\qquad$\includegraphics[height=30mm]{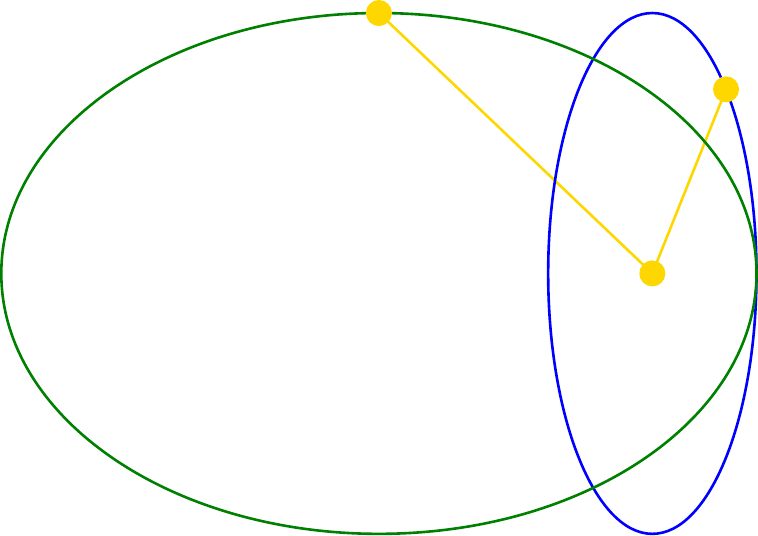}$\qquad$
\includegraphics[height=30mm]{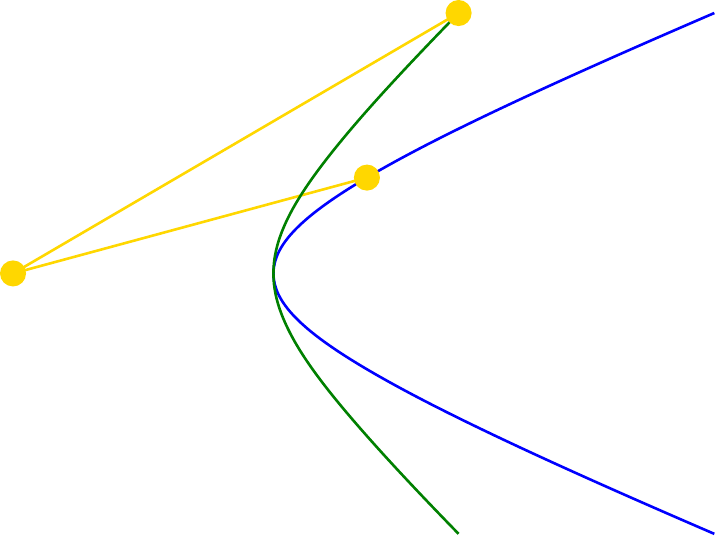}}

\nobreak
\centerline{Figure 5. The map $z\mapsto z^2$ acting on orbits with $\Re(\dot z)=0$ at $z=1$.}

\nobreak
\centerline{Two cases of Proposition {}2.6, then a case of Proposition {}2.7}

\bigskip

We can summarize MacLaurin's argument as follows. In $({}2.2)$, we wrote the square of the velocity in three ways. A fourth way is with MacLaurin's quantity ${\rm SQ}=C/\|\dot q\|$. We see that $u'^2+u^2=1/{\rm SQ}^2$. We may use $1/{\rm SQ}^2$ instead of $u'^2+u^2$. We will need to compare $1/{\rm SQ}^2$ and $1/{\rm SP}^2$. We use the similarity as we just did. We continue the proofs of \S {}2 without changing anything else.

Attempts to present the duality of central forces in the style of the {\it Principia} appear in [Nee] and [Coo]. That MacLaurin did it already in 1742 was completely ignored until N.\ Guicciardini pointed to the first author the relevant passages in MacLaurin, in January 2014. It is questionable if these passages were ever cited. We present a timeline on this subject at \S {}10.

\bigskip

{\bf {}4. The Jacobi-Maupertuis metric tensor}

We consider Lagrangian systems on a configuration space ${\cal M}$. A tangent vector $v\in \T_q {\cal M}$ at a configuration $q\in{\cal M}$ is called a {\it velocity}. The Lagrangian is always assumed to be of the form $T+U$ where $U: {\cal M}\to\R$ is a smooth function on the configuration space, called the {\it force function}, and $T: \T{\cal M}\to \R$ is a smooth function on the tangent space of ${\cal M}$, called the {\it kinetic energy}. We always assume that $T$ is, at each configuration $q$, a {\it non-degenerate quadratic function} of the velocity $v\in \T_q{\cal M}$. In other words, $T$ defines a {\it metric tensor} on the manifold ${\cal M}$, i.e., a field of non-degenerate bilinear symmetric forms. This is a pseudo-Riemannian structure on ${\cal M}$. We mainly consider the case of a Riemannian structure, for which the tensor is positive definite. To distinguish this case, we call a {\it quadratic Lagrangian} a Lagrangian $T+U$ where $T$ is quadratic and non-degenerate, and a {\it natural Lagrangian} the special case of a quadratic Lagrangian for which $T>0$ if $v\neq 0$. Under this non-degeneracy assumption, a vector field on $\T{\cal M}$ is associated to the Lagrangian, through the system of Lagrange's equations (see [Al2], \S 8.3, about the degenerate case). A solution of this well-known system of ordinary differential equations will be called a solution of the Lagrangian. Also, a Hamiltonian function and a Hamiltonian vector field is defined on $\T^*{\cal M}$, and we will speak in the same way of a solution of the Hamiltonian. Recall that the function $T-U$ is called the {\it energy} and is constant along a solution of the Lagrangian $T+U$. Here is the well-known statement which introduces the {\it Jacobi-Maupertuis metric tensor}.

{\bf {}4.1. Theorem.} Let $\I\subset\R$ be an open interval, ${\cal M}$ a configuration space. Any solution $\I\to {\cal M}$, $t\mapsto q$ of the quadratic Lagrangian $T+U$ with zero energy and nonzero force function (i.e.\ $T=U\neq 0$) is a reparametrized solution of the quadratic Lagrangian $UT$.

{\bf Proof.} At each configuration $q$ the map $L: \T_q{\cal M}\to\T^*_q{\cal M}$, $\dot q\mapsto p=\partial T/\partial \dot q$ is invertible. By using the inverse map $L^{-1}$ we define the composed function ${\cal T}: \T^*{\cal M}\to \R$, $(q,p)\mapsto (q,\dot q)\mapsto T$. A solution $t\mapsto (q,\dot q)$ of the first Lagrangian $T+U$ corresponds to a solution $t\mapsto(q,p)$ of the Hamiltonian ${\cal T}-U$. Note that the conditions $T-U=0$ and $U\neq 0$ imply $\dot q\neq 0$. Consequently equilibria, which are also critical points of the Hamiltonian, are excluded.

The second Lagrangian $UT$ is a quadratic Lagrangian on the domains where $U\neq 0$. It defines the invertible map 
$\T_q{\cal M}\to\T^*_q{\cal M}$, $\dot q\mapsto \tilde p=\partial (UT)/\partial\dot q=Up$. The second Hamiltonian is the composed function $(q,\tilde p)\mapsto (q,p)\mapsto (q,\dot q)\mapsto UT$. Here the map $(q,p)\mapsto (q,\dot q)$ is the same as above, associated to $L^{-1}$. So this Hamiltonian is the composition $$(q,\tilde p)\longmapsto (q,p) \longmapsto U(q){\cal T}(q,p)=U(q){\cal T}\Bigl(q,\frac{\tilde p}{U(q)}\Bigr)=\frac{{\cal T}(q,\tilde p)}{U(q)},$$ since ${\cal T}$ is homogeneous of degree 2 in the variable $p$, as the composition of the homogeneous maps $L^{-1}$ and $T$. The hypersurface ${\cal T}=U$ is a common level set in $\T^*{\cal M}$ of the Hamiltonian ${\cal T}-U$ and of the Hamiltonian ${\cal T}/U$. A solution $t\mapsto (q,p)$ of zero energy of the Hamiltonian ${\cal T}-U$ gives by a change of time $\tau\mapsto t$ a solution $\tau\mapsto t\mapsto (q,p)$ of energy 1 of the Hamiltonian ${\cal T}/U$, since both solutions follow the Cauchy characteristics of the hypersurface ${\cal T}=U$. By simply forgetting the variable $p$, we have the solution $t\mapsto q$ of the Lagrangian $T+U$ and the solution
$\tau\mapsto t\mapsto q$ of the Lagrangian $UT$. QED

{\bf {}4.2. Remark.} Let $g$ be the metric tensor associated to the kinetic energy $T$. A solution of the Lagrangian $UT$ is a geodesic of the metric tensor $Ug$, called the {\it Jacobi-Maupertuis metric tensor}. The traditional statement identifies an unparametrized solution of the Lagrangian $T+U$ of energy $h$ to an unparametrized geodesic of the metric tensor $(U+h)g$. We get this statement by changing $U$ into $U+h$.

{\bf {}4.3. Remark.} We gave the statement in terms of Lagrangians and the proof in terms of Hamiltonians. A proof in terms of Lagrangians, as well as the traditional deduction from the Jacobi-Maupertuis variational principle, would be longer (compare [Rou], \S 524 and p.\ 408, and the cited references to Darboux and Painlev{\'e}, [Da3], \S 571, [Pa1], p.\ 236). We used in the proof a remarkable Hamiltonian property that will now be enunciated. As in Definition {}2.3 the ratio of velocities or of momenta is computed by dividing the second quantity by the first one.

{\bf {}4.4. Proposition.} Consider the two solutions compared in Theorem {}4.1. If the second one has energy $UT=1$, then the ratio of velocities is $1/U$, and the ratio of Hamiltonian momenta is $1$.

{\bf Proof.} The previous proof gives the same $p$ for both solutions. In other words, the ratio of momenta is 1.
The respective velocities are
$L^{-1}(p)$ and $L^{-1}(p/U)$. In other words, the ratio of velocities is $1/U$. QED

{\bf {}4.5. Remark.} Choose any function $\phi: {\cal M}\to \R^+$ and consider the Lagrangian $T/\phi+\phi U$. Clearly, a solution such that $T=\phi^2 U$ is a reparametrized solution of the Lagrangian $UT$ and consequently, a reparametrization of a solution such that $T=U$ of the Lagrangian $T+U$. Larmor [Lar] and Routh [Rou], \S 628--635, present interesting examples of this.
These conformal transformations are distinct from the Darboux inversion, which we will discuss in \S {}6. The Darboux inversion maps all the solutions of positive energy of a system onto the solution of only one other system (compare Remark {}2.5). In Larmor or Routh, the image system changes when the energy changes.

{\bf {}4.6. Remark.} We stated Theorem {}4.1 for quadratic Lagrangians, and our proof indeed only uses the homogeneity and the non-degeneracy of $T$. But we will insist on the natural Lagrangians. For them, $U<0$ is excluded by the hypothesis $T-U=0$. Call ${\cal M}^+$, ${\cal M}^0$ and ${\cal M}^-$ the subsets of ${\cal M}$ defined respectively by the conditions $U>0$, $U=0$ and $U<0$. The domain ${\cal M}^+$ is called the {\it Hill region}, while ${\cal M}^0$ is called the {\it zero velocity hypersurface}. The proof of Theorem {}4.1 also proves the reciprocal statement: any non-equilibrium solution in ${\cal M}^+$ of the Lagrangian $UT$ is a reparametrized solution with zero energy of the Lagrangian $T+U$. But the Lagrangian $UT$ also defines solutions in ${\cal M}^-$, where the solutions of the Lagrangian $T+U$ have a non-zero energy.
Any non-equilibrium solution in ${\cal M}^-$ of the Lagrangian $UT$ is a reparametrized solution with zero energy of the Lagrangian $T-U$.

This is a first relation between the Lagrangians $T+U$ and $T-U$. A second one, concerning the algebraic orbits and the closed orbits, may be found in [Koe]. We will now explain a third one.

\bigskip

{\bf May a solution cross the zero velocity hypersurface?} Consider a natural Lagrangian $T+U$. It defines the above partition ${\cal M}={\cal M}^+\cup{\cal M}^0\cup{\cal M}^-$ of the configuration space. Theorem {}4.1 compares it to the Lagrangian $UT$. As this second Lagrangian vanishes on ${\cal M}^0$, its solutions are no longer defined. The velocity tends to infinity when $U\to 0$. In contrast, the solutions of the Lagrangian $T+U$ having zero energy are well-defined, but they do not cross ${\cal M}^0$. An initial condition with zero energy on this hypersurface has zero velocity. It comes from ${\cal M}^+$ and continues in ${\cal M}^+$. This is a {\it brake solution} which follows the same path in the future and in the past. Only the solutions with positive energy can cross ${\cal M}^0$. It is interesting however to compare the brake solutions when we change the sign of the force function $U$.

{\bf {}4.7. Theorem.} A solution of the real analytic Lagrangian $T+U$ starting with zero velocity from a point with $U=0$ follows the analytic continuation of the curve drawn by the solution of the Lagrangian $T-U$ with the same initial condition.

{\bf Proof.} Let us start at $t=0$ from the given point $\alpha$ with $U=0$. As the solution is the same if we change $t\to -t$, the Taylor expansion of the position $q$ in an arbitrary chart is $q(t)=\alpha+\beta t^2+\gamma t^4+\cdots$, where $q$, $\alpha$, $\beta$, $\gamma,\dots$ are vectors of real coordinates. We set $t=is$ where $i=\sqrt{-1}$. Then $q(is)=Q(s)=\alpha-\beta s^2+\gamma s^4+\cdots$ is a real curve parametrized by $s$. 
Let us write the equations of motion in terms of the Christoffel symbols $\Gamma_{jkl}$ of the Levi-Civita connection of the metric tensor $g$ associated to $T$. Let $f=\nabla U$ be the force field. The differential equation for $q=(q_1,\dots,q_n)$ is $\ddot q_j+\sum_{kl}\Gamma_{jkl}\dot q_k\dot q_l=f_j(q)$. After the change to time $s$, the left-hand side changes sign while $f$ remains the same. So, the equation of motion for $Q(s)$ is the same as the equation of motion for $q(t)$, except that the force field is $-f$ instead of $f$. Thus $Q(s)$ is the solution of the Lagrangian $T-U$. We set $u=t^2$ and now $u\mapsto \alpha+\beta u+\gamma u^2+\cdots$ explicitly gives the announced analytic continuation. QED

 \bigskip
\centerline{\includegraphics[width=60mm]{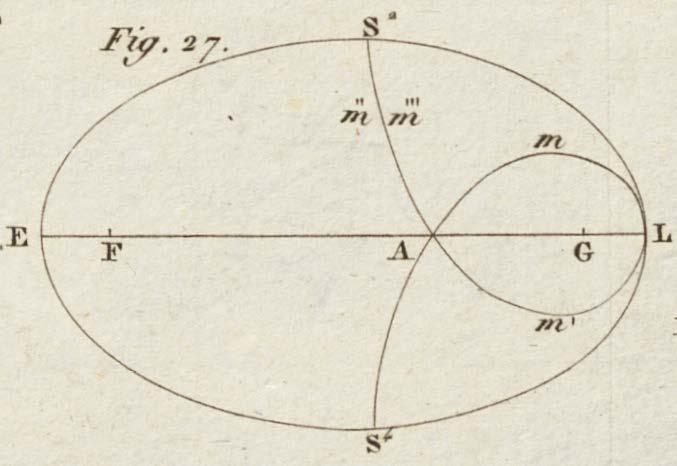}}

\nobreak
\centerline{Figure 6. A brake orbit for a body attracted by two fixed centers F and G}

\nobreak
\centerline{Figure of [Leg2], source: ETH-Bibliothek}
\bigskip

Legendre discovered in the two fixed centers problem what were maybe the first brake trajectories. In [Leg], \S 209 or [Leg2], \S 509, he writes: ``Au reste, une courbe alg\'ebrique ne pouvant pas \^etre termin\'ee brusquement en ${\rm S}^2$ et ${\rm S}^4$, les branches ${\rm AS}^2$, ${\rm AS}^4$ ont sans doute une continuation...'' As we saw, this continuation is a trajectory of the integrable Lagrangian obtained by changing the sign of the force function. The idea of associating this change of sign to the imaginary time is explained for example in [App] or [Koe].

\bigskip

{\bf {}5. Smooth isometric coverings of the Jacobi-Maupertuis surface}

Given a natural Lagrangian on a 2-dimensional configuration space ${\cal M}$, the {\it Jacobi-Maupertuis surface} is ${\cal M}$ endowed with the Jacobi-Maupertuis metric tensor.

Here we show that a regularization of the dynamics of a Kepler problem on a given energy manifold is obtained by passing to the Jacobi-Maupertuis surface and taking the isometric double covering of it. The same process is also successful for the Newton-Lobachevsky force function on a surface of constant curvature. The same process, with an $n$-covering instead of double, works for McGehee's force functions. These observations show the close relation of this process with the material of \S {}2. We will confirm this relation in \S {}6.

{\bf {}5.1. The Jacobi-Maupertuis cone of a homogeneous force function.} We start with the configuration space ${\cal M}=\R^2\setminus \{\O\}$ with the metric tensor $dx^2+dy^2=dr^2+r^2d\theta^2$ and the force function $U=r^{2k}$. In other words, the kinetic energy is $T=(\dot x^2+\dot y^2)/2$ and the Lagrangian is $T+U$. Here we consider the zero energy solutions, satisfying $T-U=0$. The Jacobi-Maupertuis metric tensor is $r^{2k}(dr^2+r^2d\theta^2)$. The variable $\rho=r^{k+1}/(k+1)$, or, if $k=-1$, $\rho=\log r$, satisfies $d\rho=r^kdr$. The Jacobi-Maupertuis metric tensor becomes $d\rho^2+(k+1)^2\rho^2d\theta^2$, or, if $k=-1$, $d\rho^2+d\theta^2$. The perimeter of the circle of radius $\rho$ is $2\pi\rho |k+1|$, or, if $k=-1$, $2\pi$. The surface is a cone, or, if $k=-1$, a cylinder. The cone cannot be embedded in a Euclidean space if $|k+1|>1$, since the perimeter increases with the radius $\rho$ faster than it does in the plane. When $k$ increases from $-1$ to $0$, the cone opens up. At $0$, the cone is a plane. As $k>0$, it fails to be embedded. The same occurs when $k$ decreases from $-1$.

{\bf {}5.2. Remark.} The Gaussian curvature of the cone and of the cylinder is zero. That it should be zero in general follows from this argument by Goursat [Gou] and Darboux [Da2]. If the metric tensor $dx^2+dy^2$ of the Euclidean plane $\O xy$ is multiplied by a conformal factor $\lambda$, the Gaussian curvature $\kappa$ is given by 
$$\kappa=-\frac{1}{2\lambda}\Bigl(\frac{\partial^2 \log\lambda}{\partial x^2}+\frac{\partial^2 \log\lambda}{\partial y^2}\Bigr)\eqno({}5.1)$$ according to [Lio], p.\ 295. Here the conformal factor is $\lambda=r^{2k}$. Let $\O xy$ be the complex plane $\C$. We have $\Delta(\log r^{2k})=\Delta (k \log (z\bar z))=k\Delta (\log z+\log\bar z)=0$ since the real part of a holomorphic function is harmonic. Consequently $\kappa=0$.

If we set $\omega=(k+1)\theta$, the metric tensor becomes Euclidean. The map $(r,\theta)\mapsto (\rho,\omega)$ is the map $z\mapsto z^{k+1}/(k+1)$, the same as in \S {}2 except for the new denominator $k+1$. Here the Jacobi-Maupertuis surface only concerns the zero energy solutions. We will see in \S {}6 how Darboux removed this restriction. But let us see how to remove it by a generalization which includes the non-zero energy and the non-zero constant curvature.

A complete surface ${\cal M}$ of constant curvature $\kappa$ may be embedded in $\R^3=\Omega xyz$ endowed with the metric tensor $dx^2+dy^2+\kappa^{-1}dz^2$ as a connected component of the surface $\kappa(x^2+y^2)+z^2=1$. The division by $\kappa$ does not give a singularity at $\kappa=0$. If moreover a point $\O\in{\cal M}$ is chosen, the embedding may be chosen in such a way that $\O=(0,0,1)$. We call ${\cal M}_\O$ the part of the surface with $z>0$. If $\kappa>0$, ${\cal M}_\O$ is a hemisphere centered at $\O$. If $\kappa=0$, ${\cal M}_\O$ is the Euclidean plane. If $\kappa<0$, ${\cal M}_\O$ is a pseudo-sphere.

{\bf {}5.3. Definition.} Consider a surface ${\cal M}$ of constant curvature $\kappa$ and a point $\O\in {\cal M}$. Consider the central projection of center $\Omega$ and of image the plane $\O xy$, as defined by the above embedding. We call the {\it projected radius} the function ${\cal M}_\O\to \R$, $q\mapsto r$ where $r\geq 0$ is the distance from $\O$ to the central projection of $q$.

If ${\cal M}$ is a sphere of radius 1, i.e., $\kappa=1$, the projected radius $r$ is the tangent of the angle $\O q$.
It is infinite on the equator, i.e., the boundary of ${\cal M}_\O$. Note that $1/r$, the cotangent, has a smooth continuation on and beyond the equator. The following Theorem shows in a new way the regularity property of the {\it McGehee exponents} that we have met in Theorem {}2.9.

{\bf {}5.4. Theorem.} Consider, on a surface ${\cal M}_\O$ of constant curvature with an origin $\O$, the force function $r^{2k}$, where $r$ is the projected radius, and where $k=1/(l+1)-1$ for some $l\in\N$. The isometric $l+1$-fold covering, ramified at $\O$, of the Jacobi-Maupertuis surface at energy $h$, extends as an analytic surface at $\O$. Furthermore, the function $r^{-2k}$ is analytic at $\O$ under the same condition.

{\bf Proof.} The tangent plane at $\O$ being considered as a local chart $\O xy$ of the surface through the central projection, the function $r$ satisfies $r^2=x^2+y^2$ and the metric tensor is $(1+\kappa r^2)^{-2}\bigl(dx^2+dy^2+\kappa(xdy-ydx)^2\bigr)$ which is also $$\frac{dr^2}{(1+\kappa r^2)^2}+\frac{r^2 d\theta^2}{1+\kappa r^2}.\eqno({}5.2)$$ We multiply by $r^{2k}(1+hr^{-2k})$ to get the Jacobi-Maupertuis metric tensor at energy $h$. We use the above variables $\rho=r^{k+1}/(k+1)$, $\omega=(k+1)\theta$. The latter relation defines a locally isometric surface. The metric tensor is now
$$(1+hr^{-2k})\Bigl(\frac{d\rho^2}{(1+\kappa r^2)^2}+\frac{\rho^2 d\omega^2}{1+\kappa r^2}\Bigr).$$
Let $X=\rho\cos\omega$ and $Y=\rho\sin\omega$. We reduce to the same denominator. The quantity $d\rho^2+\rho^2d\omega^2=dX^2+dY^2$ extends as an analytic metric tensor at $\O$.
As $r^2=(1+k)^2r^{-2k}\rho^2$, the other term $\kappa r^2(XdY-YdX)^2/\rho^2$, the denominator and the first factor are analytic at $\O$ if $r^{-2k}$ is analytic. For a McGehee exponent $k$, $-k/(k+1)=l\in\N$ and $r^{-2k}=(k+1)^{2l}\rho^{2l}$, which is analytic. QED

{\bf {}5.5. Jacobi-Maupertuis surface for the Newton-Lobachevsky force function.}
We first consider the Newtonian force function $U=1/r$ in the plane. What is the geometry of the Jacobi-Maupertuis surface with given energy $h_0$? What is the geometry of the double covering?
Next, we consider the Newton-Lobachevsky force function, which is the known analogue of $U=1/r$ on a surface of constant curvature. This is again $1/r$, but where $r$ is the projected radius defined in {}5.3. We will present in Theorem {}5.8 the particular case with the simplest geometry.

{\bf {}5.6. Theorem (Moeckel).} The Jacobi-Maupertuis surface of the Kepler problem at energy $h_0<0$ has a positive curvature, has a vertex with angle $\pi/3$ at the collision and another vertex at the zero velocity curve. Near the latter vertex, where the radius is $r_0=-1/h_0$, the surface cannot be embedded as a surface of revolution in the Euclidean $\R^3$. The isometric embedding finishes at the radius $3r_0/4$. The generatrix is parametrized by the radius $r\in[0,3r_0/4]$ in term of elliptic functions and elliptic integrals.

\bigskip

\centerline{\includegraphics[width=40mm]{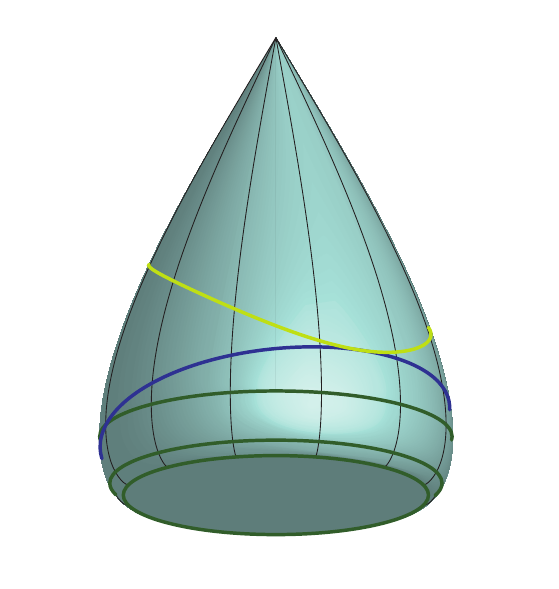}\includegraphics[width=65mm]{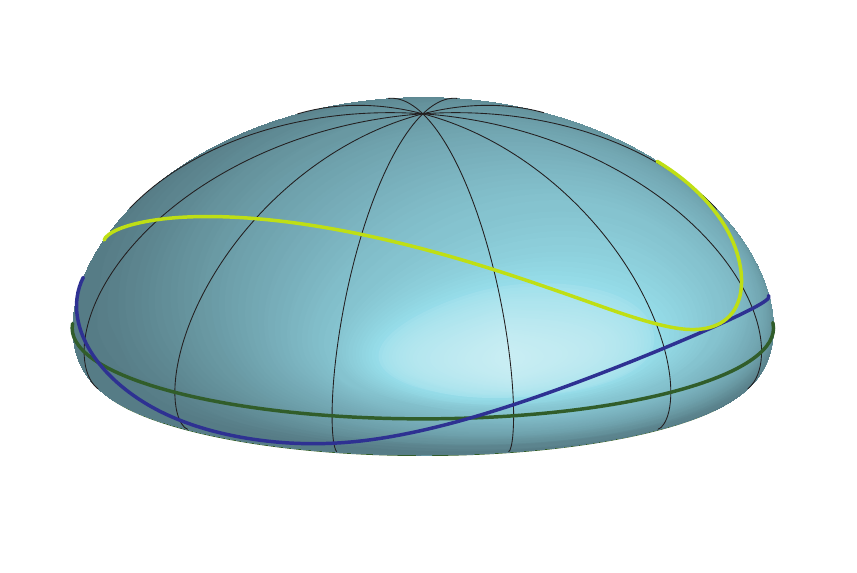}}

\nobreak
\centerline{Figure 7. The Jacobi-Maupertuis surface of Kepler, $h_0<0$, and its covering}
\bigskip

The obstruction to embedding is the same as for the cone: The perimeter of a parallel grows too quickly. Moeckel [Moe] also presents embeddings in Minkowski space, as well as the less problematic case $h_0>0$. But let us try the double covering. We propose the following statement for comparison.

{\bf {}5.7. Theorem.} The isometric double covering of the Jacobi-Maupertuis surface of the Kepler problem at energy $h_0<0$ has a positive curvature, extends as an analytic surface at the collision and has a vertex at the zero velocity curve. Near the vertex, where the radius is $r_0=-1/h_0$, the surface cannot be embedded as a surface of revolution in the Euclidean $\R^3$. The isometric embedding finishes at the radius $2r_0/3$. The generatrix is parametrized by the radius $r\in[0,2r_0/3]$ in term of trigonometric functions and arcs.

{\bf Proof.} The smoothness at the collision is the particular case of Theorem {}5.4 for $k=-1/2$ and zero curvature. The impossibility of embedding is Theorem 1 of [Moe] applied to the double covering. The formula for the generatrix is computed as in [Moe], where we double the value of $f$. We get consequently
$g'(r)^2=(2-3r)/(1-r)$ instead of $g'(r)^2=(3-4r)/(4r-4r^2)$.
Integrating $g'(r)dr$, we get a trigonometric arc instead of an elliptic integral, since the degree of the denominator is one instead of two. QED

The next result will appear as equivalent to the result by Nersessian and Pogosy\-an in [NeP] as soon as we will present the Darboux inverse. It will also appear in a simple way in {}9.6.

{\bf {}5.8. Theorem.} Consider a surface of constant curvature $\kappa<0$
endowed with the Newton-Lobachevsky force function $r^{-1}$, where $r$ is the projected radius defined in {}5.3. Set $c=\pm\sqrt{-\kappa}$. The isometric double covering of the Jacobi-Maupertuis surface at energy $h=c$ extends at the collision as an analytic surface of constant curvature $-c/2$. It is a hemisphere if we choose $c<0$ and a pseudo-sphere if we choose $c>0$.

{\bf Proof.} We make $\kappa=-c^2$ in $({}5.2)$ and multiply by the Jacobi-Maupertuis conformal factor $r^{-1}+c$. We change $r$ into $\rho$ defined by $4\rho^{-2}=r^{-1}+c$, which gives $8\rho^{-3}d\rho=r^{-2}dr$. We have $1-c^2r^2=r^2(r^{-1}+c)(r^{-1}-c)=4r^2\rho^{-2}(4\rho^{-2}-2c)=16r^2\rho^{-4}\Phi$ where $\Phi=1-c\rho^2/2$. The Jacobi-Maupertuis metric tensor is
$\Phi^{-2}d\rho^2+(4\Phi)^{-1}d\theta^2$. Setting $\theta=2\omega$, it becomes 
$\Phi^{-2}d\rho^2+\Phi^{-1}d\omega^2$.
Comparing with formula $({}5.2)$, we see that this metric tensor has constant Gaussian curvature $-c/2$. The image of the central projection is the disk $r^{-1}>|c|$. If $c<0$, this is exactly the domain $4\rho^{-2}=r^{-1}+c>0$. So, this is a hemisphere. QED

\bigskip

{\bf {}6. The Darboux inverses}

In \S {}4 a natural Lagrangian $T+U$, restricted to an energy level set, is transformed into a geodesic flow on the Jacobi-Maupertuis manifold. In this section we endow this manifold with the force function $1/U$ or $-1/U$. This gives two new natural Lagrangians with same solutions, up to reparametrization, as $T+U$, on some {\it open set of the phase space}, and not just on an energy surface (compare Remark {}2.5). We call them the Darboux inverses. The process is an involution: we get a duality between natural Lagrangians which, when combined with some local isometries, includes the MacLaurin duality of central forces described in \S {}2.

 Darboux explained these facts in two pages in 1889. These pages were forgotten and the level of generality they provide was never reached again. Before oblivion, the pages were indeed continued under the name of ``Transformation des {\'e}quations de la dynamique'', notably by R.\ Liouville, Painlev\'e and Levi-Civita. But this theory was also forgotten, which is not astonishing since the basic examples of transformation were forgotten. They were later rediscovered, notably by Faure, Higgs and McGehee (see the timeline in \S {}10). Of this old theory of transformations with change of time, the only part which is still studied concerns the geodesics. Levi-Civita [Lev] is often cited, for example in [MaT]. Levi-Civita later published an important application of the Darboux inversion, the regularization of the 3-body problem. He described Darboux's result and the corresponding change of time in [Le4].

We will identify reparametrized solutions of the natural Lagrangian $T+U$ in the three generic cases corresponding to the partition of the configuration space ${\cal M}={\cal M}^+\cup{\cal M}^0\cup{\cal M}^-$ (respectively, $U>0$, $U=0$, $U<0$) and to the sign of the energy.

{\bf {}6.1. Theorem (Darboux, 1889).} A solution in ${\cal M}^+$, of arbitrary positive energy $h$, of the natural Lagrangian $T+U$ is a reparametrized solution of energy $1/h$ of the natural Lagrangian $UT+1/U$.

{\bf First proof (Darboux).} By Theorem {}4.1 and Remark {}4.2 the first solution is a reparametrized solution of the Lagrangian $(U+h)T$ while the second solution is a reparametrized solution of the Lagrangian $(1/U+1/h)UT$. But these are the same since
$$(U+h)T=\Bigl(\frac{1}{U}+\frac{1}{h}\Bigr)hUT\eqno({}6.1)$$ and the factor $h$ is irrelevant. QED

{\bf Second proof.} Consider
the Lagrangian $\sqrt{h}(UT+1/U+1/h)$, which has the same solutions as the Lagrangian $UT+1/U+1/h$. With the notation ${\cal T}$ in the proof of Theorem {}4.1, the associated Hamiltonian is
${\cal T}/(U\sqrt{h})-(1/U+1/h)\sqrt{h}$. The zero energy level has equation ${\cal T}=U+h$. Now the Lagrangian $T+U+h$ defines the Hamiltonian ${\cal T}-U-h$ and the same zero energy level. We conclude by the Cauchy characteristics as in the proof of Theorem {}4.1. QED

{\bf {}6.2. Proposition.} Consider the two solutions compared in Theorem {}6.1. The ratio of velocities is $1/(U\sqrt{h})$. The ratio of momenta is $1/\sqrt{h}$.

{\bf Proof.} According to the second proof, at a configuration $q$ the momentum $p$ is the same for the Lagrangians $T+U+h$ and $\sqrt{h}(UT+1/U+1/h)$. We have $p=\partial (\sqrt{h}UT)/\partial \dot q$ while the momentum for the Lagrangian $UT+1/U+1/h$ is $\partial (UT)/\partial \dot q=p/\sqrt{h}$. So the ratio of momenta is $1/\sqrt{h}$. We divide this ratio by $U$ to get the ratio of velocities as in Proposition {}4.4. QED

{\bf {}6.3. Remark.} In the above statements $T=U+h>0$, which excludes the possibility of a zero velocity hypersurface $U+h=0$ and of an equilibrium. In the next statements $h$ and $U$ have different signs. We should consider the special case of an equilibrium. Also, in the first proof, the intermediate Jacobi-Maupertuis geodesic flow has a singularity on the zero velocity hypersurface $U+h=0$. In the second proof, we avoid this intermediate problem and the singularity, if we take care of studying separately the equilibria.

{\bf {}6.4. Theorem.} The solutions in ${\cal M}^+$ of arbitrary negative energy $h$ of the natural Lagrangian $T+U$ are reparametrized solutions of energy $-1/h$ of the natural Lagrangian $UT-1/U$. For a non-equilibrium solution the ratio of velocities is $1/(U\sqrt{-h})$ and the ratio of momenta is $1/\sqrt{-h}$.

{\bf Proof.} The solutions of $T+U$ satisfy the inequality $U+h\geq 0$ which is $1/U+1/h\leq 0$ since $h<0$ and $U>0$. According to Remark {}4.6, the non-equilibrium solutions of $(1/U+1/h)UT$ are reparametrized solutions of the Lagrangian $UT-1/U-1/h$. We adapt the previous proofs to these new signs. The equilibria are the critical points of $U$ in the configuration space, with zero velocity. Consequently, they are also the critical points of $1/U$ with zero velocity. The equilibria are the critical points of the Hamiltonian in the phase space: if we exclude them, the level set of the Hamiltonian is smooth and we can apply the Cauchy characteristic argument of the second proof. QED

{\bf {}6.5. Remark.} The hypothesis $h<0$ induces the minus sign in the formula $TU-1/U$. In the first proof, this is explained by Remark {}4.6. In the second proof, this is explained by the introduction of the factor $\sqrt{h}$, which we should replace by $\sqrt{-h}$.

{\bf {}6.6. Theorem.} The solutions in ${\cal M}^-$ of the natural Lagrangian $T+U$ have positive energy $h=T-U$. They are reparametrized solutions of energy $-1/h$ of the natural Lagrangian $-UT-1/U$. For a non-equilibrium solution the ratio of velocities is $-1/(U\sqrt{h})$ and the ratio of momenta is $1/\sqrt{h}$.

{\bf Proof.} We may use the second proof of Theorem {}6.1 and change in the end the Lagrangian $UT+1/U$ into $-UT-1/U$ in order to get a natural Lagrangian. This also changes the sign of the energy. We consider the equilibria separately as in the proof of Theorem {}6.4. QED

When passing to a reparametrized solution the sign of the force function and the sign of the energy are exchanged. In Theorem {}6.1, the exchange is $(+,+)\mapsto (+,+)$. In Theorem {}6.4, it is $(+,-)\mapsto (-,+)$. In Theorem {}6.6, it is $(-,+)\mapsto (+,-)$. The pair $(-,-)$ is impossible.
We recognize the rule of signs of Propositions {}2.7 and {}2.8. In all the cases we observe during the computation an exchange of the force function and of the energy. This exchange also explains the minus sign of Remark {}6.5.

We may check accordingly that the Theorems define involutive steps. For a positive energy in ${\cal M}^+$ we reparametrize twice in a row according to Theorem {}6.1 and Proposition {}6.2. We come back to the initial solution of the initial Lagrangian. Starting with negative energy in ${\cal M}^+$ Theorem {}6.4 gives the transformed Lagrangian $UT-1/U$. Thus ${\cal M}^+$ is renamed ${\cal M}^-$ and Theorem {}6.6 applies. We come back to the initial solution of the initial Lagrangian.

The above discussion of signs with three cases is not so easy. We propose to simplify the rule by allowing the non-natural quadratic Lagrangians, as suggested in the proof of Theorem {}6.6. We reduce the discussion to two cases.

{\bf {}6.7. Definition.} In the domain of the configuration space where $U\neq 0$, we define the {\it first Darboux inverse} of the quadratic Lagrangian $T+U$ as the quadratic Lagrangian $UT+ 1/U$. We define the {\it second Darboux inverse} of $T+U$ as the quadratic Lagrangian $-UT+ 1/U$.

{\bf {}6.8. Theorem.} The solutions with force function $U\neq 0$ and energy $h\neq 0$ of the quadratic Lagrangian $T+U$ are reparametrized solutions of energy $1/h$ of the first Darboux inverse if $h>0$, of the second Darboux inverse if $h<0$.

{\bf Proof.} We use for example the second proof of Theorem {}6.1 if $h>0$.
We compare the equilibria as in the proof of Theorem {}6.4. We apply the Cauchy characteristic argument after removing the equilibria. In the case $h<0$, we replace $\sqrt h$ by $\sqrt{-h}$. QED

\bigskip

{\bf Example of the force function on the Kepler cone.}
We consider the Kepler problem in the plane, i.e., the natural Lagrangian $T+U=(\dot x^2+\dot y^2)/2+1/r$ on the configuration space $\R^2\setminus \{(0,0)\}$. 
Theorems {}6.1 and {}6.4 propose to endow the {\it Kepler cone} with the force function $1/U=r$ if $h>0$, with $-r$ if $h<0$. 
Montgomery [Mon] and Moeckel [Moe] have recently published about the Jacobi-Maupertuis surface of the Kepler problem, and called it the Kepler cone. This is an ordinary cone with angle $\pi/3$ at the apex, as we have seen at the beginning of \S {}5. Theorem {}5.4 claims that after an isometric double covering, the surface becomes smooth at the apex and the function $r$ becomes smooth on this covering. 
These are simple facts. We will repeat the formulas. The Jacobi-Maupertuis metric tensor is $r^{-1}(dr^2+r^2 d\theta^2)$. We set $\rho=2r^{1/2}$ and it becomes $d\rho^2+\rho^2d\theta^2/4=d\rho^2+\rho^2d\omega^2$ if $\omega=\theta/2$. If $\theta$ is the polar angle, this is the Kepler cone. If $\omega$ is the polar angle, this is the Euclidean metric tensor in the plane.
{\it The Darboux inversion introduces the force function $1/U=r=\rho^2/4$ on the Kepler cone. We pass to the double covering and get the repulsive Hooke force function in the plane}. If the Keplerian energy is negative, we have the attracting Hooke force function in the plane. In brief, we replace the $z\mapsto \sqrt{z}$ complex map of \S {}2 by two steps: $(i)$ pass to the Jacobi-Maupertuis surface with the Darboux force function and $(ii)$ take the isometric double covering. We find in particular that the new surface and the new force function are smooth. Consequently the dynamics is smooth. We propose the following generalizations.

{\bf {}6.9. Proposition.} If $k=1/(l+1)-1$ is a McGehee exponent, consider the isometric covering ${\cal C}$ of the Jacobi-Maupertuis surface defined in Theorem {}5.4. Darboux's function $(r^{2k}+h)^{-1}$ is analytic at the origin $\O$ of ${\cal C}$. In the special case $k=-1/2$ of a Newton-Lobachevsky law of force on a surface of constant negative curvature $\kappa$, where we choose $h=\pm\sqrt {-\kappa}$, we have $(r^{-1}+h)^{-1}=\rho^2/4$, where $\rho$ is the projected radius on the surface ${\cal C}$, which according to Theorem {}5.8 has constant curvature.

{\bf Proof.} According to Theorem {}5.4, $r^{-2k}$ is an analytic function at $\O$. Consequently $(r^{2k}+h)^{-1}=r^{-2k}(1+hr^{-2k})^{-1}$ is also analytic at $\O$. Its value $\rho^2/4$ in the special case is given in the proof of Theorem {}5.8. QED

The square of the projected radius is the known Hooke force function in constant curvature ([Ser], p.\ 205, see also [AlZ]). As announced before Theorem {}5.8, we have now recovered the result of [NeP]. We get after two steps what they get with the map $z\mapsto z^2$. We may also compute Darboux's force function for the examples listed at the beginning of \S {}5. We had $U=r^{2k}$. We express $1/U$ as a function of $\rho$, which gives, up to a factor, $\rho^{-2k/(k+1)}$. This is again the dual force law of \S {}2. Darboux's presentation gives an interesting answer in the special case $k=-1$. The dual force function is defined on the ordinary cylinder obtained in the beginning of \S {}5. The force function is $1/U=r^2=e^{2\rho}$ where $\rho$ is the height on the vertical cylinder.

\bigskip

{\bf {}7. Back to 1877 for another insight of Darboux}

Darboux raised in 1877 the following question: which are the {\it central Lagrangians} having an open set in their phase space filled up with periodic solutions? By a central Lagrangian
we mean a system defined by a surface of revolution together with a force function which is constant on the parallels.
Bertrand had previously asked the question of the periodic solutions for a central force in the plane. As Bertrand and Darboux we will only discuss the case of a 2-dimensional configuration space, since the same question in the 3-dimensional case is easily reduced to the 2-dimensional case.

 Darboux treated separately (in 1877, 1886, 1894) the case where the force function is constant, i.e., the geodesic flow on a surface of revolution, and his study was continued with the examples by Tannery in 1892 and Zoll in 1903. Darboux gave in 1877 a complete classification in the case where the force function is analytic and non-constant. He introduced an interesting class of central Lagrangians.
 
We will show in \S {}9 that this class is invariant by the Darboux inversion, which Darboux introduced in 1889. {\it This invariance explains the open sets of periodic solutions}. A Darboux inversion sends locally a solution onto a solution, and consequently a closed solution onto a closed solution, except if the image meets some boundary. All the systems that Darboux proposed in 1877 are obtained through his 1889 inversion from the Kepler problem on a surface of constant curvature. Consequently all the considered surfaces of revolution are Jacobi-Maupertuis surfaces of the Kepler problem on a surface of constant curvature. Even if Darboux slightly reworked his 1877 material in 1886, he apparently never related it with his inversion.

Darboux indeed introduced two classes, which will both appear to be inversion invariant. The {\it general class} of central Lagrangians has a force function $u$ which Darboux takes as the radial coordinate on the surface with metric tensor
$$ds^2=\frac{\mu^2}{2}\frac{\varpi''(u)}{\varpi^2(u)}du^2+\frac{d\omega^2}{\varpi(u)}.\eqno({}7.1)$$
Here $\omega$ is the angular coordinate and $\mu\neq 0$ is a real number. The definition of the function $\varpi(u)$ is clear: Since it is the denominator of the second term in the right-hand side, it satisfies
$$\varpi(u)=\frac{1}{r^2},\eqno({}7.2)$$
where $2\pi r$ is the perimeter of the circle with force function $u$. In the case of an immersed surface of revolution, $r$ is the distance to the axis of symmetry. According to $({}7.1)$, $\varpi$ should be positive and convex, with a non-zero second derivative.

The choice of the force function $u$ as a radial coordinate simplifies the formulas (compare [San]).
It may be noticed that the radial Clairaut-Binet variable for the Kepler problem is also the force function, and that the formula for the Darboux inverse also suggests this choice. Darboux proves that the condition
$$3\varpi''\varpi''''-4\varpi'''^2=0\eqno({}7.3)$$
expresses that a periodic solution on a parallel has a neighborhood filled up with periodic solutions. We will call the {\it rational class} the subclass of the general class which solves this equation. This is $$\varpi(u)=\frac{a_0 u^2+a_1 u+a_2}{a_3 u+a_4}\eqno({}7.4)$$
where $a_0,\dots, a_4$ are real constants. Darboux excludes the case where $\varpi$ is an affine function of $u$, since the metric tensor $({}7.1)$ would be degenerate. There remains two cases. The first case has $a_3=0$. Then we may choose $a_4=1$ and
$$\varpi(u)=a_0u^2+a_1u +a_2,\quad \hbox{with } a_0\neq 0.\eqno({}7.5)$$ 
A change $u\mapsto u+\gamma$ with $\gamma\in \R$ allows to assume $a_1=0$. The second case is $a_3\neq 0$. We can normalize $a_3=1$ and choose $a_4=0$ after a change $u\mapsto u+\gamma$, giving
$$\varpi(u)=a_0u+a_1+\frac{a_2}{u},\quad \hbox{with } a_2\neq 0.\eqno({}7.6)$$
If we need the general expression, we may introduce $u_0\in \R$ and write
$$\varpi(u)=a_0(u-u_0)+a_1+\frac{a_2}{u-u_0},\quad \hbox{with } a_2\neq 0.\eqno({}7.7)$$
We will now state these results more accurately.

{\bf {}7.1. Definition.} A {\it central Lagrangian} is a natural Lagrangian $T+U$ on the annulus $\A=\I\times \s$, where $\s$ is the circle, $\I\subset\R$ is an open interval, which is invariant by the action of the circle on $\A$, $(r,\omega)\mapsto (r,\omega+c)$ where $c\in\R$, $r\in\I$ is a radial coordinate and $\omega\in\R/2\pi\Z$ an angular coordinate. A {\it parallel solution} or relative equilibrium is a solution of a central Lagrangian which remains on a parallel, i.e., where the radial coordinate remains constant.

{\bf {}7.2. Remark.} Here $T$ defines the metric tensor giving to the annulus $\A$ the geometry of a surface of revolution, while $U$ is the force function generating a force field which is central if we think of $\A$ as an annulus in the plane. We do not assume that the surface of revolution is embedded in $\R^3$. The main counterexample is the Darboux inverse of the Kepler problem. We have seen Moeckel's description {}5.6 of its surface of revolution, which cannot be embedded. Darboux does not really assume the embedding since his variable $r$ may as well be interpreted as the perimeter divided by $2\pi$.

{\bf {}7.3. Remark.} For simplicity the configuration space is defined as an open annulus. It does not include any apex, even if in interesting cases the surface of revolution extends smoothly at an apex (see Figure 7).

{\bf {}7.4. Definition.} A {\it Darboux Lagrangian} with parameter $\mu>0$ is a central Lagrangian $T+u$ for which the force function $u$ is a coordinate on $\I$, and for which the kinetic energy $T$ is the metric tensor $({}7.1)$ divided by $2dt^2$. Here $\varpi: \I\to \R$ is required to be a rational function of the form $({}7.4)$, satisfying $\varpi>0$ and $\varpi''>0$ on $\I$, and such that $\I$ is maximal for these properties, i.e., is not strictly included in an interval on which $\varpi>0$ and $\varpi''>0$.

{\bf {}7.5. Theorem (Darboux).} An analytic central Lagrangian which possesses a parallel solution having a neighborhood in the phase space filled up with periodic solutions is the restriction of a Darboux Lagrangian to an annulus, with parameter $\mu\in\Q$.

{\bf Proof.} See [Da1] or [DD2]. Note that our hypothesis implies that there is an {\it arbitrarily small} invariant neighborhood of the parallel solution filled up with periodic solutions. This is needed in Darboux's argument. QED

{\bf {}7.6. Remark.} The graph of $u\mapsto \varpi$
shows where are the parallel solutions. Let us try to see this intuitively. Along such a solution, the centrifugal acceleration tends to increase the radius $r$, while the force tends to increase the force function $u$. They balance each other if $r$ decreases when $u$ increases, and if moreover we choose the convenient angular velocity. As $\varpi=1/r^2$, {\it there is a parallel solution} at each $u$ such that $\varpi'(u)>0$. The hypothesis of the Theorem concerns a single parallel solution, but a neighborhood for this solution is also a neighborhood for the neighboring parallel solutions. So Darboux condition $({}7.3)$ is true on an interval. It extends analytically. The rational expression $({}7.4)$ is this analytic extension. In particular $\varpi$ is a single-valued function of $u$, while $u$ is not always a single-valued function of $\varpi$. This is another justification for Darboux's choice of $u$ as the radial coordinate. {\it The analytic extension is not stopped at a $u_0$ such that $\varpi'(u_0)=0$, called an equator.} Equators do happen (see the examples in Figure 7). In [ZKF] the study of the equators is extended to the non-analytic case. We may remark that the convexity condition $\varpi''>0$ shows that there is at most one equator in the domain of a Darboux Lagrangian.
Also, a Darboux Lagrangian $T+u$ possesses at each $u$ such that $\varpi'(u)>0$
a parallel solution with a neighborhood filled up with periodic solutions. At a $u$ such that $\varpi'(u)<0$, the Lagrangian $T-u$ possesses a parallel solution with this property.

{\bf {}7.7. Remark.} For a given parameter $\mu$, the same function $({}7.4)$ defines one or two domains for a Darboux Lagrangian, provided we assume $a_0>0$ in case $({}7.5)$ in order to get the convexity. Indeed, in case $({}7.5)$ normalized by the condition $a_1=0$, there are two intervals if $a_2\leq 0$ and one if $a_2>0$. Each of these intervals gives a Darboux Lagrangian. In case $({}7.6)$, $u$ has the sign of $a_2$ on the allowed intervals. An allowed interval is bounded by $u=0$. There may be a second one, as we will see in \S {}8.

{\bf {}7.8. The role of the parameter $\mu$ and the smooth extension.} If we choose $r$ as the radial coordinate then the metric tensor $({}7.1)$ becomes
$$ds^2=2\mu^2\frac{\varpi''\varpi}{\varpi'^2}dr^2+r^2d\omega^2.\eqno({}7.8)$$
If $2\mu^2\varpi''\varpi=\varpi'^2$, this is the Euclidean metric tensor expressed in the polar coordinates $(r,\omega)$. If the surface is smooth in the neighborhood of a point with $r=0$, then $r$ is equivalent to the geodesic distance $s$ to this point, and consequently $2\mu^2\varpi''\varpi/\varpi'^{2}\to 1$ as $r\to 0$. In the case $({}7.5)$ with $a_0>0$, when $u\to \pm\infty$, this limit is $\mu^2$. Thus {\it the metric tensor may be smoothly extendable in this case only if $\mu^2= 1$.} That it is smoothly extendable under this condition will be proved in {}8.2. In the case $({}7.6)$ with $a_2>0$, when $u\to 0^+$, this limit is $4\mu^2$. Thus, {\it the metric tensor may be smoothly extendable in this case only if $\mu^2= 1/4$.} That it is smoothly extendable under this condition will be reduced by Theorem {}9.3 to Theorem {}5.4. The last case with $r\to 0$ is $({}7.6)$ when $u\to \pm\infty$. In this case this limit is zero and {\it the metric tensor is not smoothly extendable}. The main example is again Theorem {}5.6.

In 1886 Darboux added to his 1877 paper the formula for the $z$ coordinate of the embedding of the surface of revolution in $\R^3$:
$$dz^2=\frac{2\mu^2\varpi(u)\varpi''(u)-\varpi'^2(u)}{4\varpi^3(u)}du^2.\eqno({}7.9)$$
Interestingly, the factor is $-(1/\varpi)''/8$ if $\mu=1/2$. Darboux also observed the following: If another annulus with another central Lagrangian is defined as $({}7.1)$ with the letters $(\mu_1,\varpi_1,u_1,\omega_1)$, if there is an $\alpha\neq 0$ such that $\varpi_1=\alpha^2\varpi$, $\mu_1=\alpha\mu$, then the map $(u,\omega)\mapsto (u_1,\omega_1)= (u,\alpha\omega)$ preserves the metric tensor $ds^2$. {\it This map is a local isometry.} We may use it to select the values $\mu=1$ or $\mu=1/2$ which give a smooth apex.

\bigskip
\centerline{\includegraphics[width=60mm]{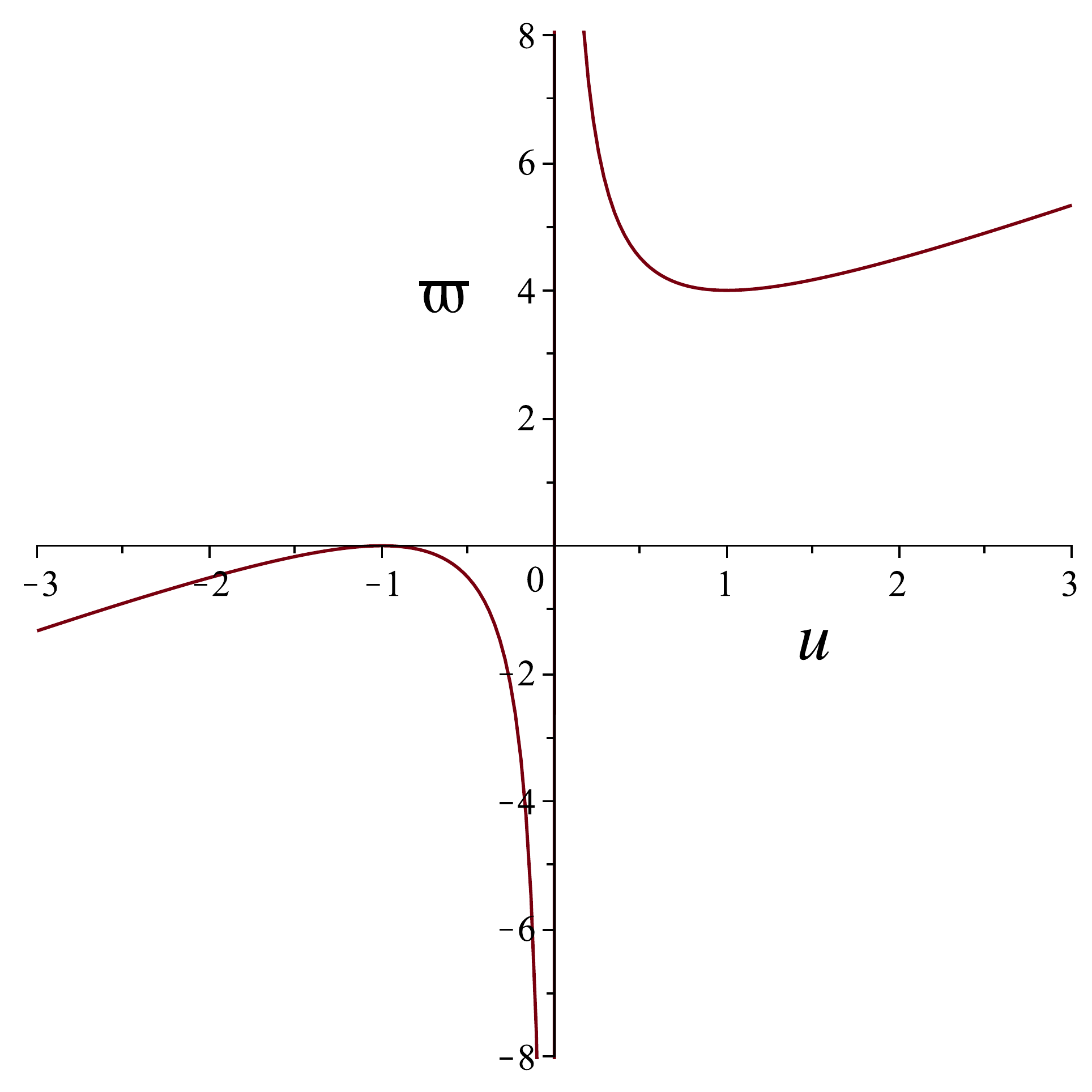}}

\nobreak
\centerline{Figure 8. The graph of $\varpi$ for the Darboux Lagrangians of Figure 7}
\bigskip

{\bf {}7.9. The non positive definite continuation.} The metric tensor $({}7.1)$ changes signature as $\varpi$ or $\varpi''$ change sign. The graph of a rational $\varpi$ typically displays open intervals in $u$ where this metric tensor is indefinite or negative definite. We do not study the indefinite cases, even if they are of course of great interest (see [Zag]). We observe that the negative definite case reduces to the positive definite case by a simple change $\varpi\mapsto -\varpi$. Nevertheless we have seen that Theorem {}6.8 is simpler to state if we accept the negative definite Lagrangians.

\bigskip

{\bf {}8. The curvature of a Darboux Lagrangian}

The Gaussian curvature $\kappa$ of a surface of revolution with metric tensor
$$ds^2=E(u)du^2+G(u)d\omega^2$$
may be expressed by merely writing the famous formula given by Gauss in his {\it Disquisitiones generales circa superficies curvas}, [Gau] \S 11, in this particular case:
$$4E^2G^2\kappa=EG'^2+GE'G'-2EGG''$$
or
$$\kappa=-\frac{1}{4G'}\Bigl(\frac{G'^2}{EG}\Bigr)'.\eqno({}8.1)$$
For the general class $({}7.1)$, this is
$$\kappa=\frac{\varpi^2}{2\varpi'\mu^2}\Bigl(\frac{\varpi'^2}{\varpi''\varpi}\Bigr)'.\eqno({}8.2)$$
For the rational class, this gives in the first case $({}7.5)$ a constant curvature
$$\kappa=\frac{4a_0a_2-a_1^2}{4\mu^2a_0},\eqno({}8.3)$$
and in the second case $({}7.6)$ a polynomial of degree 3 in $u$
$$\kappa=\frac{2a_0^2u^3+3a_0a_1u^2+6a_0a_2u+a_1a_2}{4\mu^2a_2}.\eqno({}8.4)$$
The curvature $({}8.4)$ is constant if and only if $a_0=0$. If $a_0\neq 0$, the discriminant of $\kappa$ in the variable $u$ is $-27a_0^3(a_1^2-4a_0a_2)^2/64 a_2^3\mu^8$. This formula singles out the particular case $a_1^2-4a_0a_2=0$, where we have
$$a_2=\frac{a_1^2}{4a_0},\quad\varpi(u)=\frac{(2a_0u+a_1)^2}{4a_0 u},\quad \kappa=\frac{(2a_0u+a_1)^3}{4\mu^2 a_1^2}.\eqno({}8.5)$$
As we see, the curvature does not change sign on an allowed branch, where $\varpi>0$.
Let us see the general case $({}7.6)$ on a convex positive branch, i.e., with $a_2/u>0$ and $\varpi(u)>0$. We compute
$$\kappa'=\frac{3a_0(a_0u^2+a_1u+a_2)}{2\mu^2a_2}=\frac{3a_0u\varpi}{2\mu^2a_2}.\eqno({}8.6)$$ Remarkably, there is a factor $\varpi$. On the allowed branches, $\kappa'$ has the sign of $a_0$, which proves the following

{\bf {}8.1. Theorem ([ZKF]).} On a surface of revolution $\A$ of a Darboux Lagrangian the Gaussian curvature $\kappa$ is monotone. Consequently it may change sign on at most one parallel.

We further remark the unexpected factor $\varpi$ in the general class, since formula $({}8.2)$ gives $$\kappa'=-\frac{\varpi}{2\mu^2}\Bigl(\frac{\varpi'\varpi'''}{\varpi''^2}\Bigr)'.\eqno({}8.7)$$
The discussion of all the cases with non-constant curvature is rather simple. We can restrict the discussion to the case $a_2>0$ in formula $({}7.6)$, where only $u>0$ is allowed by the convexity condition.

If $a_0<0$, the discriminant of $\kappa$ is positive, $\kappa$ has three roots, while $\varpi=(a_0 u^2+a_1u+a_2)/u$ is positive on a unique interval, bounded below by $u=0$. If $a_1>0$, $\kappa$ has two variations: two roots are positive. Between them, there is the local minimum of $\kappa$, which corresponds to a root of $\varpi$, and which is the upper bound of the unique interval where $\varpi>0$. The curvature changes sign once on $\A$. If $a_1\leq 0$, $\kappa$ has one variation, thus one positive root. The zero of $\varpi$ is smaller, and the curvature is negative on $\A$.

If $a_0>0$, the discriminant of $\kappa$ is negative, $\kappa$ has only one root, and $\varpi>0$ for all $u>0$, except if $\kappa$ has two positive critical points, in which case $\varpi\leq 0$ between them. In this case, the root of $\kappa$ cannot be in the excluded interval, since $\kappa$ would have 3 roots. If $a_1\geq 0$, the root of $\kappa$ is non-positive, $\varpi$ has no positive root, and the unique Darboux Lagrangian has a positive curvature. If $a_1<0$, the root of $\kappa$ is positive. If $a_1^2-4a_0a_2<0$, there is only one Darboux Lagrangian. Its curvature changes sign once. If $a_1^2-4a_0a_2>0$, there are two Darboux Lagrangians. We need another special property of the curvature polynomial, suggested by $({}8.5)$, to determine the sign of the curvature on each of them. Let us call the mid-value of a polynomial of degree 3 the value at the inflection point. The mid-value of $({}8.4)$ is computed at $u=-a_1/2a_0$. It is
$$\kappa=\frac{a_1(a_1^2-4a_0a_2)}{8a_0a_2\mu^2}.\eqno({}8.8)$$
This value is negative in the present case. Consequently the curvature of the first Darboux Lagrangian, which corresponds to the interval bounded below by zero, is negative. The curvature of the second Darboux Lagrangian changes sign once. The discussion of the cases is now complete.

{\bf {}8.2. Cases with constant curvature.} As we said, the case $({}7.5)$ may be reduced to the subcase with $a_1=0$ by a translation of $u$, giving $\varpi=a_0u^2+a_2$. We may choose $\mu=1$ to get the smooth extension as $u\to\infty$. Then the curvature is the constant $\kappa=a_2$. Using $\varpi=1/r^2$ as in $({}7.8)$
$$ds^2=\frac{\varpi}{a_0 u^2}dr^2+r^2d\omega^2=\frac{1}{1-\kappa r^2}dr^2+r^2d\omega^2.\eqno({}8.9)$$
This is the pull-back by orthogonal projection on a plane of the metric tensor of a sphere in a Euclidean space or a pseudo-sphere in a Minkowskian space. The convexity of $\varpi$ gives $a_0>0$. We set $a_0=1/m^2$. The force function is $u= m(r^{-2}-\kappa)^{1/2}$. If $\kappa=0$, this is the Kepler problem with mass $m$. The graph of $u\mapsto \varpi$ is a parabola cut in two branches by the condition $\varpi>0$. The branch with $u>0$ is the attractive Kepler problem, the branch with $u<0$ the repulsive Kepler problem.
If $\kappa>0$ there is only one branch, the whole parabola. The radius of the sphere is $\rho=\kappa^{-1/2}$. Then $r=\rho\sin\theta$ where $\theta$ is the angle from the North pole, and $u=m (\rho\tan\theta)^{-1}$ which is the Serret force function, i.e., the Newton-Lobachevsky force function in positive curvature. If $\kappa<0$ we set $\rho=(-\kappa)^{-1/2}$, $r=\rho\sinh\tau$, and get $u=m (\rho\tanh \tau)^{-1}$, which is the Killing force function, i.e., the Newton-Lobachevsky force function in negative curvature.

In the case $({}7.6)$ with $a_0=0$, $\varpi=a_1+a_2/u$ and the curvature $\kappa=a_1/4\mu^2$ is constant. We should choose $\mu=1/2$ to get the smooth extension when $u\to 0$. Then $\kappa=a_1$ and
$$ds^2=\frac{u\varpi}{a_2}dr^2+r^2d\omega^2=\frac{1}{1-\kappa r^2}dr^2+r^2d\omega^2.$$
This is the metric tensor $({}8.9)$.
We may set $a_2=m$ as this parameter is comparable to the gravitational mass in the previous example. The force function is $u=m/(r^{-2}-\kappa)$. If $\kappa=0$ this is the Hooke repeller in the plane if $m$ and thus $u$ are positive, and the Hooke attractor if $m$ and thus $u$ are negative. If $\kappa=1/\rho^2>0$ we set $r=\rho \sin\theta$ and get $u=m(\rho\tan\theta)^2$, which is the hemispherical Hooke attractor or repeller. If $\kappa<0$ the convex branch is cut by the condition $\varpi>0$. Let $\kappa=-1/\rho^2$, $r=\rho\sinh \tau$, $u=m(\rho\tanh\tau)^2$. This is the Hooke attractor or repeller on the pseudo-sphere. We may check that, whatever the curvature, $u$ is proportional to the square of the projected radius defined in {}5.3.

\bigskip
{\bf {}9. The Darboux inverse of a Darboux Lagrangian.}

We will show that Darboux's conclusion about the central Lagrangians with an open set filled up with closed orbits, which we explained in \S {}7, may be expressed without formulas. The idea is extremely simple. It is enough to combine this conclusion with the subsequent result by Darboux explained in \S {}6.

{\bf {}9.1. Lemma.} Let $\mu\in\R$ be a parameter, $\I\subset \R^+$ or $\I\subset \R^-$ an open interval, $\varpi:\I\to\R$ a non-vanishing smooth function. On the annulus $\A=\I\times \s$, where $\s$ is a circle, we have
$$u\Bigl(\frac{\mu^2}{2}\frac{\varpi''(u)}{\varpi^2(u)}du^2+\frac{d\omega^2}{\varpi(u)}\Bigr)=\frac{\mu^2}{2}\frac{\rho''(v)}{\rho^2(v)}dv^2+\frac{d\omega^2}{\rho(v)},\eqno({}9.1)$$
 where $\omega\in\R/2\pi\Z$ is an angular coordinate on $\s$, where $u$ is a coordinate on $\I$, where $v=1/u$ and $\rho(v)=v\varpi(1/v)$.

{\bf Proof.} We have the remarkable identity $v^3\rho''(v)=\varpi''(1/v)$. This identity is presented at any order of differentiation in [Com] p.\ 170 or [ComE] p.\ 161 and is credited to [Ha2]. QED

This lemma shows that {\it the general class of central Lagrangians}, with metric tensor $({}7.1)$, {\it is invariant by Darboux's inversion}. Indeed, the Darboux inverses of the Lagrangian $T+u$ are $u T+1/u$ and $-u T+1/u$. The metric tensor associated to $u T$ is given by $({}9.1)$ and $v$ is the new force function.

{\bf {}9.2. Theorem.} If $T+u$ is a Darboux Lagrangian (see Definition {}7.4) then the Darboux inverses $u T+1/u$ and $-u T+1/u$ are Darboux Lagrangians. If $T$ corresponds to the fraction $\varpi$ of $({}7.4)$, then $u T$ as a function of $v=1/u$ corresponds to the fraction
$v\mapsto\rho(v)=v\varpi(1/v)$.

{\bf {}9.3. Theorem.} Any Darboux Lagrangian is, up to the addition of a constant, either a Newton-Lobachevsky force function on an annulus of constant curvature or a Darboux inverse of such a system.

{\bf Proof.} There are two classes of Darboux Lagrangians, corresponding to the formulas $({}7.5)$ and $({}7.6)$. The Newton-Lobachevsky systems correspond to $({}7.5)$. Their Darboux inverses correspond to $v \varpi (1/v)=a_0/v+a_1+a_2v$ with $a_0\neq 0$, which is the second class. QED

{\bf {}9.4. Remark.} Adding a constant to the force function does not change the dynamics but does change the Darboux inverse. For example in the Kepler problem, if we take $u=1/r$, the surface is an ordinary cone (see the end of \S {}6), while if $u=-1+1/r$, the surface is represented in Figure 7. Note that a Darboux inverse of a Newton-Lobachevsky system has a force function which is zero at an apex. If we apply the Darboux inversion again, we come back to Newton-Lobachevsky. If we first add a constant and then apply the inversion, we stay in the class $({}7.6)$. Note also that $({}8.5)$ is a Darboux inverse of the Kepler problem in the plane.

{\bf {}9.5. Example of the Kepler problem.} Here we consider the simplest case, the force function $u=m/r$, where $m$ is the mass parameter. It is the case of $({}7.1)$ with $\mu=1$ and $\varpi(u)=(u/m)^2$. The attractive case is $u>0$. The first Darboux inverse, which concerns the positive energy, corresponds to $\mu=1$ and $\rho(v)=v\varpi(1/v)=(m^2v)^{-1}$. According to {}7.8 the smooth isometric covering
corresponds to $\mu=1/2$ and $\rho(v)=(4m^2v)^{-1}$. On the corresponding domain $v>0$ this is the repulsive Hooke problem with positive energy. The ``mass'' coefficient is $(4m^2)^{-1}$.

The second Darboux inverse, which concerns the attractive Kepler problem in negative energy, corresponds to $\mu=1$ and $\rho(v)=-(m^2v)^{-1}$. The smooth isometric covering
corresponds to $\mu=1/2$ and $\rho(v)=-(4m^2v)^{-1}$. Formula $({}7.1)$ gives a negative definite $ds^2$. We may change this sign without changing the dynamics by considering the natural Lagrangian $-ds^2/(2dt^2)-v$. But now the force function is $w=-v$. Both changes of sign change $\rho$ into $-(4m^2w)^{-1}$, which we consider with $w<0$. This is the attractive Hooke problem with necessarily positive energy.

The repulsive Kepler problem corresponds to the same $\varpi(u)$ but with $u<0$. The energy is positive, so we consider the first Darboux inverse, which again corresponds to $\mu=1$ and $\rho(v)=(m^2v)^{-1}$. The smooth isometric covering is $\mu=1/2$ and $\rho(v)=(4m^2v)^{-1}$.
Again this gives a negative definite $ds^2$ which becomes natural after the same two changes of sign, which change $\rho$ into $(4m^2 w)^{-1}$, to be considered with $w>0$. This is the repulsive Hooke problem with negative energy. We got the three cases illustrated in Figure 5.

{\bf {}9.6. Example of the constant curvature cases.} Let us see how to rediscover the result of Nersessian and Pogosyan described in {}5.8 and {}6.9.
We have noticed in \S {}8 that there are surfaces of constant curvature in the second class. We can reach them by Darboux inversion from the case $a_2=0$ of $({}7.5)$, namely, $\varpi=a_0u^2+a_1u
$. If $\mu=1$, which gives the surface with smooth apex, then $\kappa=-a_1^2/4a_0$ according to $({}8.3)$. According to {}8.2, $a_0=1/m^2$ where $m$ is the mass. {\it We should take the Darboux inverse of a Newton-Lobachevsky force function with mass $m$ on a surface of constant negative curvature $\kappa$.} But according to the condition $a_2=0$, there is a choice of constant in the force function $u$. The position should tend to infinity when $u\to 0$.

 Now the first Darboux inverse corresponds to $\mu=1$ and $\rho(v)=v\varpi(1/v)=a_0/v+a_1$. The locally isometric surface with smooth apex has $\mu=1/2$ and $\rho(v)=a_0/4v+a_1/4$. The curvature is $\kappa=a_1/4$ and the ``mass'' is $a_0/4$ according to {}8.2. This confirms the data in {}5.8 and {}6.9, where we used the normalization $m=1$. We have obtained the Hooke problem in constant curvature and thus confirmed the result of Nersessian and Pogosyan.

We will furthermore discuss the possible cases: attraction, repulsion, negative or positive curvature, negative or positive energy. Let us see the allowed intervals for the variable $u$. The hypothesis $a_0>0$ gives a convex $\varpi(u)=a_0u^2+a_1u$. We first assume that $a_1>0$. Then $u\in \R^+$ and $u\in (-\infty, -a_1/a_0)$ are allowed, since these intervals give a positive definite $ds^2$ in $({}7.1)$.
The first is the Newton-Lobachevsky attraction. The above formulas for the first Darboux inverse concern the positive energy case.
They give the Hooke problem in constant positive curvature $a_1/4$ with a positive ``mass'' $a_0/4$, which means a repulsion. The energy is positive. The second Darboux inverse concerns the negative energy case. Using the same rule as in the flat case, $\rho$ becomes $-(4m^2w)^{-1}+a_1/4$, with $w=-v<0$. This is the attractive Hooke problem in positive curvature, with positive energy.

Now if $u\in (-\infty, -a_1/a_0)$, we have the Newton-Lobachevsky repulsion with positive energy $T-u$. We take the first Darboux inverse, which corresponds to $\rho(v)=a_0/4v+a_1/4$ with $v\in (-a_0/a_1,0)$. The $ds^2$ is negative definite. We pass to $-ds^2/(2dt^2)-v$ to get a natural Lagrangian. The force function is $w=-v\in (0,a_0/a_1)$, and the formula for $\rho$ is $a_0/4w-a_1/4$. This is a repulsive Hooke problem in negative curvature $-a_1/4$ with negative energy.

We got three cases of the Hooke problem in non-zero constant curvature. The hypothesis $a_1<0$ will give the other three. The domain $u<0$ has only positive energy solutions. The first Darboux inverse gives a repulsive Hooke problem in positive curvature with negative energy. The case $u\in (-a_1/a_0, +\infty)$ has both signs of the energy. The first Darboux inverse has $v\in (0, -a_0/a_1)$. It is a repulsive Hooke problem in negative curvature with positive energy. The second Darboux inverse after changes of sign has $w\in (a_0/a_1,0)$. It is an attractive Hooke problem in negative curvature with positive energy.

Let us summarize this discussion. Following Nersessian and Pogosyan we consider the generalization to a non-zero constant curvature surface of the planar map $z\mapsto z^2$. The discussion of the possible signs of the energy and the ``mass'' parameter is the same as in Figure 5, where the three possibilities were obtained in the planar case. But we can furthermore choose the sign of the curvature in the Hooke problem, and this will not affect the curvature in the resulting Newton-Lobachevsky problem. The resulting curvature is always a negative constant. This leads to a kind of paradox, since this map is essentially a Darboux inversion, which is involutive and consequently one-to-one. The paradox is solved by observing that the change of sign of the curvature will change the Newton-Lobachevsky force function by adding a constant to it. The force function vanishes at infinity if and only if we start from the Hooke problem in positive curvature. This addition of a constant will be significant when applying the Darboux inversion a second time, or in other words, when applying the Jacobi-Maupertuis conformal factor.

\bigskip

{\bf {}10. Timeline on $z\mapsto z^2$}

1720. In his {\it Geometria organica} [Mac], MacLaurin gives a Proposition XXII on the sinusoidal spirals as solutions of central forces problems. Cooper [Coo] relates it to the duality of central forces.

1742. Colin MacLaurin in his {\it Treatise of fluxions} [Ma2], \S 451, states the duality of homogeneous central forces, which he proves at \S 875.

1817. Legendre confirms by several arguments the extended motion after collision with a center in the two fixed centers problem: ``il s'approche du centre G et s'en \'eloigne ensuite, en revenant sur ses pas, comme s'il y avait en G un obstacle in\'ebranlable qui ne perm{\^\i}t pas au corps de passer outre, et qui le f{\^\i}t rejaillir en sens contraire avec la m\^eme vitesse.'' ([Leg], \S 228 or [Leg2], \S 528)

1878. Halphen [Hal] remarks the invariance by homography of both classes he had found as the solutions of the question: what are the central force fields in the plane having only conic sections as solutions? This question is due to Bertrand (see [Be1], [Al1]).

1889. Goursat [Gou] rediscovers MacLaurin results via Hamilton-Jacobi equation and conformal maps.

1889. Darboux [Da2] reacts to Goursat work and extends it to the curved and multidimensional case by introducing what we call the Darboux inversion.

1890. Appell [Ap1] extends Halphen's remark on invariance by homography to any force field on an affine space.

1891. Appell [Ap2] remarks that the central projection sends the orbits of the Kepler problem onto orbits of Serret's extension to Kepler problem on the sphere.

1894. Painlev\'e publishes [Pai] about ``transformation des {\'e}quations de la Dynamique''. Starting with the examples by Darboux and Appell, he asks for all the transformations sending a system onto another system with a change of time.
He analyses the previous works on page 17.

1896. Painlev\'e [Pa2] announces a complete classification for the case with two degrees of freedom, with Appell's and Darboux's examples as the two main cases.

1900. In [RLC], Ricci and Levi-Civita, continuing [Lev], cite Painlev\'e's question as a motivation for the development of the ``calcul diff\'erentiel absolu'', which is also called covariant differentiation or Ricci calculus. They do not recall the basic examples.

1903. First announcements by Levi-Civita about the regularization of binary collisions in the restricted 3-body problem (see [Le1]).

1904. Levi-Civita [Le2] announces the use of $z\mapsto z^2$ as a regularizing transformation in the restricted 3-body problem in the plane.

1906. Levi-Civita publishes this regularization in Acta Mathematica [Le3].

1907. First publication by Sundman about the collisions in the 3-body problem [Sun].

1909. Sundman [Su2] treats the extension of the solutions after a binary collision in the 3-body problem.

1911. Bohlin [Boh] presents the transformation $z\mapsto z^2$ as part of a new method of integration the Kepler problem. He does not give any reference.

1912. Sundman ([Su3], p.\ 141) wrote: ``On voit donc que les orbites des corps $P_0$ et $P_1$ pr\'esenteront chacune un point de rebroussement au point o{\`u} ces corps viennent se choquer. Au contraire l'orbite du corps $P_2$ restera continue dans le voisinage de l'instant $t_1$ du choc. Les orbites que d\'ecrivent ainsi les corps apr{\`e}s le choc se rattachent du reste d'une fa{\c c}on continue aux orbites qu'on obtient en faisant varier d'une fa{\c c}on continue les valeurs initiales des coordonn\'ees et des composantes des vitesses de ces corps de fa{\c c}on {\`a} faire passer les corps tout pr{\`e}s les uns des autres sans se heurter.'' (The last sentence is added compared to [Su2], p.\ 13. The distinction between extension and regularization is carefully discussed in [McG], with examples and counter-examples. See Theorem {}2.9.)

1913. Kasner [Kas] explains and develops the ``Projective Transformations'' by Appell and the ``Conformal transformations'' with the example of the MacLaurin duality. The references for the latter material are to ``Larmor, Goursat and Darboux'', through the discussion in [Rou], \S 628--635. Indeed Routh does not cite Goursat. The reader would only find the reference to [Da2] by detecting the reference [Pa1] at the last page of [Rou], and by going on reading the lithography [Pa1] in French, where Painlev{\'e} does not choose the simplest exposition of [Da2]. Larmor and Routh do not explain the Darboux inversion, but only material related to Remark {}4.5.

1915. Motivated by Sundman's regularization of the 3-body problem in dimension 3, Levi-Civita publishes a series of five notes in Rendiconti dei Lincei where he extends his 1906 regularization from the planar restricted case to the planar general case. In the first note [Le4], he chooses the change of time associated to the Darboux inversion of the 3-body problem.

1916. Levi-Civita [Le5] presents a regularization of the 3-body problem in the 3-dimensional case, using part of the ingredients of the Darboux inversion and what he calls the parabolic osculating elements. His main formulas are reproduced in [Sie], [SiM], \S 7, (31), (32), [Mos] (2.10).

1920. Levi-Civita [Le6] reorganizes his 1915--1916 researches and publishes them in Acta Mathematica, in French.

1924. Levi-Civita [Le7] presents the same researches in a simpler way.

1941. Wintner [Win], p.\ 423, cites Bohlin [Boh] for his ``elegant method of integration''.

1946. Thomas [Tho] discusses and extends to the $n$-dimensional case the classification of transformations reviewed in Ricci and Levi-Civita (1900).

1953. Faure ([Fau], [Fa2]) remarks that the conformal transformation $z\mapsto z^2$ pulls back the Schr\"odinger equation of the planar atom into that of the planar harmonic oscillator. More generally he describes the same results as Goursat in this planar quantum mechanics context, including the MacLaurin duality. He does not cite any reference.

1964. Using spinors, Kustaanheimo [Kus] presents a regularization of some perturbed Kepler problem in dimension 3, which reduces to $z\mapsto z^2$ when restricted to a plane.

1965. Kustaanheimo and Stiefel [KSt] present the previous regularization as a transformation from the three-dimensional Kepler problem into the system of 4 isotropic harmonic oscillators in $\R^4$ subject to a constraint and with an additional circle symmetry (see [Zha]).

1979. Higgs [Hig] rediscovers Appell's central projection from the planar to the spherical Kepler problem.

1981. McGehee studies the regularization of the homogeneous force functions. He uses the definitions of regularization by Conley and Easton, but also discusses the classical works by Sundman and Levi-Civita. While citing [Le6], he only uses the map $z\mapsto z^2$ of [Le3]. He then generalizes it into $z\mapsto z^n$ where $n$ is an integer. He does not relate this extension to any previous work.

1989. Arnol'd and Vasil'ev ([ArV], [ArV2], [Arn]) present the duality of central forces, citing Bohlin (1911) and Faure (1953).

2001. Nersessian and Pogosyan [NeP] describe a conformal map from the Hooke systems on the sphere and pseudosphere to the Coulomb system on the pseudosphere. It is a composition of $z\mapsto z^2$ with stereographic projections.

\bigskip

{\bf {}11. Timeline of Bertrand's question about closed orbits}

1622. Kepler ({\it Epitome}, IV-III-III, p.\ 591, [Kep], [KepE]) wrote ``But it is unbelievable that the planet---this freedom allowed---should, after completing its return, be restored to exactly the same distance.''

1687. Newton ({\it Principia}, Book 1, Proposition 45, example 2, [New], [NewE]) wrote ``the angle VC$p$, completed in the descent of a body from the upper apsis to the lower apsis in the very nearly circular orbit which any body describes under the action of a centripetal force proportional to A$^{n-3}$, will be equal to an angle of $180/\sqrt{n}$ degrees.'' Newton considered that the ``fact that the aphelia of the planets are at rest'' shows that the law of force is in $1/r^2$, i.e., $n=1$
(General Scholium [NewE], p.\ 943).

1766. Lagrange ([Lag], p.\ 280) presented a computation in a neighborhood of a circular solution which improved that of Newton. Lagrange did at second order what Newton had done at first order. His formula solves Bertrand's question (1873) for the homogeneous laws of force.

1873. Bertrand [Ber] posed the question of closed orbits and solved it. The statement and the proof were imprecise (see [KF3], p.\ 410, [Jov]). His followers (see notably [DD2], note 14) replaced Bertrand's argument with a computation equivalent to that of Lagrange (1766).

1877. Darboux [Da1] posed the more general question on a surface, and gave two families of solutions in the case of a non-constant potential. He also gave information in the case of a constant potential, which is the case of geodesic motions on the surface.

1884. Darboux ([DD1], note 11, see [Al1]) considers central forces which do not depend only of the distance. He recalls a class of examples due to Jacobi, with same homogeneity as the Newton law, which have open sets filled up with periodic solutions, or with algebraic solutions of chosen degree.

1886. Darboux ([DD2], note 15) published a revised version of his article from 1877. He gave in footnote two references to Paul Serret. He added in the end two examples in the case of a constant potential. These are actually Jacobi-Maupertuis metric tensors of the Kepler and the harmonic oscillator problems.

1892. Tannery [Tan] used the method of Darboux developed in [DD2] to construct an explicit algebraic closed surface of revolution on which all geodesics are closed and moreover algebraic. This is also an example on which all geodesics are closed but not all of the same length. The equators are twice shorter than the other geodesics.

1894. Darboux ([Da4], pp.\ 6--9) posed the problem of closed geodesics and gave on page 9 the conclusion in the case of surfaces of revolution with equators. He concludes that there is only the sphere and surfaces locally isometric to the sphere. He forgot to mention the hypothesis that the equator should define a plane of symmetry. This hypothesis appears in the reasoning on page 8, and in the 1877 version of the same theorem.

1903. Zoll [Zol] gave an example of a surface of revolution without singularities and with an analytic contour, on which all geodesics are closed and of same length. There is a unique equator but the surface is not symmetric with respect to the equatorial plane. He cited Darboux's book of 1894 but not the references before, and he did not consider adding a potential.

1904. Pfister in his thesis [Pfi] cited Darboux 1877 and 1886 and remarked that the singular orbits of Darboux's examples are not closed.

1913. Funk [Fu1] derived a general formula for surfaces of revolutions on which all non-singular geodesics are closed. He recovered the case of the surface derived from the Jacobi metrics of Hooke and Kepler problems and explained their link. He mentioned that these were considered in [Pfi]. He also covers the example of Zoll.

1978. Besse [Bes] signs a collective book on the topic of manifolds with only closed geodesics. The works of Darboux before 1894 are not cited.

2008. Santoprete [San], without citing any work of Darboux, addressed the problem of potentials on surfaces of revolutions with closed orbits and re-established part of Darboux's results. He worked out a generalized Bertrand theorem on surfaces of constant curvature and was aware of the possibility of having surfaces of non-constant curvature with exactly one Bertrand potential of generalized Hooke type, and gave a criterion.

2009. Based on the concluding remark of Perlick [Per], Ballesteros, Enciso, Herranz and Ragnisco [BEHR], reobtained Darboux's classification of 1877. The first integrals are given in each case.

2012. Fedoseev, Kudryavtseva and Zagryadski\v{\i} [ZKF] extended Darboux's investigation to non-analytic abstract surfaces of revolutions without equators and gave a complete classification of the surfaces and potentials for which all non-singular bounded orbits are closed. This work refers to Darboux [Da1], [DD2], [Per], [BEHR] and was continued in [KF1], [KF2], [KF3], [Zag].

\bigskip

{\it Acknowledgements.} We wish to thank Julian Barbour, Alain Chenciner, Giovanni Federico Gronchi, Niccol\`o Guicciardini and Elana A.\ Kudryavtseva for pointing us very interesting references. Lei Zhao is supported by DFG 605/1-1.

\bigskip

{\bf REFERENCES}

\noindent[Al1] A. Albouy, Projective dynamics and classical gravitation, Regular and Chaotic Dynamics, 13 (2008), pp.\ 525--542

\noindent[Al2] A. Albouy, Projective dynamics and first integrals, Regular and Chaotic Dynamics, 20 (2015), pp.\ 247--276

\noindent[AlZ] A. Albouy, L. Zhao, Lambert's theorem and projective dynamics, Philosophical Transactions of the Royal Society A: mathematical, physical and engineering sciences, 377 (2019), 20180417

\noindent[App] P. Appell, Sur une interpr{\'e}tation des valeurs imaginaires du temps en M{\'e}canique, {\it Comptes Rendus Acad.\ Sci.\ Paris,} 87 (1878), pp.\ 1074--1077

\noindent[Ap1] P. Appell, De l'homographie en m{\'e}canique, {\it American Journal of Mathematics}, 12 (1890), pp.\ 103--114

\noindent[Ap2] P. Appell, Sur les lois de forces centrales faisant d{\'e}crire {\`a} leur point d'application une conique quelles que soient les conditions initiales, {\it American Journal of Mathematics}, 13 (1891), pp.\ 153--158

\noindent[ArV] V.I. Arnol'd, V.A. Vasil'ev, Newton's Principia, Read 300 Years Later, {\it Notices of the American Mathematical Society}, 36 (1989), pp.\ 1148--1154

\noindent[ArVa] V.I. Arnol'd, V.A. Vasil'ev, Addendum to Newton's Principia, Read 300 Years Later, {\it Notices of the American Mathematical Society}, 37 (1990), p.\ 144

\noindent[Arn] V.I. Arnol'd, {\it Huygens and Barrow, Newton and Hooke: Pioneers in mathematical analysis and catastrophe theory from evolvents to quasicrystals}, B\^ale, Birkh{\"a}user, 1990

\noindent[BEHR] Ballesteros, A., Enciso, A., Herranz, F.J., Ragnisco, O., Hamiltonian systems admitting a Runge-Lenz vector and an optimal extension of Bertrand's theorem to curved manifolds, {\it Communications in Mathematical Physics}, 290 (2009), pp.\ 1033--1049

\noindent[Ber] J. Bertrand, Th\'eor\`eme relatif au mouvement d'un point
attir\'e vers un centre fixe, {\it Comptes Rendus Acad.\ Sci.\ Paris} 77
(1873), pp.\ 849--853

\noindent[Be1] J. Bertrand, Sur la possibilit\'e de d\'eduire d'une
seule des lois de Kepler le principe de l'attraction, {\it Comptes Rendus Acad.\ Sci.\ Paris} 84 (1877), pp.\ 671--674

\noindent[Bes] A.L. Besse, Manifolds all of whose geodesics are closed, Ergebnisse der Mathematik und ihre Grenzgebiete 93, Springer-Verlag, Berlin-Heidelberg-New York (1978)

\noindent[Boh] K. Bohlin, Note sur le probl{\`e}me des deux corps et sur une int{\'e}gration nouvelle dans le probl{\`e}me des trois corps, Bulletin Astronomique, 28 (1911), pp.\ 113--119

\noindent[BoM] A.V. Borisov, I.S. Mamaev, Relations between integrable systems in plane and curved spaces, {\it Celestial Mechanics and Dynamical Astronomy} 99 (2007), pp.\ 253--260

\noindent[Che] Chernikov, N. A., The Kepler problem in the Lobachevsky space and its solution, Acta physica Polonica, B 23.2 (1992), pp.\ 115--122

\noindent[Com] L. Comtet, Analyse Combinatoire, Tome 1, Presses Universitaires de France (1970)

\noindent[ComE] L. Comtet, Advanced Combinatorics, Reidel, Dordrecht (1974)

\noindent[Coo] J. B. Cooper, On the relevance of the differential expressions $f^2+f'^2$, $f+f''$ and $ff''-f'^2$ for the geometrical and mechanical properties of curves, arXiv:1102.1579 (2011)

\noindent[Da1] G. Darboux, {\'E}tude d'une question relative au mouvement d'un point sur une surface de r{\'e}volution, {\it Bulletin de la Soci{\'e}t{\'e} Math{\'e}matique de France}, Tome 5 (1877), pp. 100--113

\noindent[Da2] G. Darboux, Remarque sur la Communication pr{\'e}c{\'e}dente, {\it Comptes Rendus Acad.\ Sci.\ Paris}, 108 (1889), pp.\ 449--450

\noindent[Da3] G. Darboux, Le\c cons sur la th\'eorie g\'en\'erale des surfaces et les applications g\'eom\'etri\-ques du calcul infinit\'esimal, Partie 2, Gauthier-Villars 1889, 1915

\noindent[Da4] G. Darboux, Le\c cons sur la th\'eorie g\'en\'erale des surfaces et les applications g\'eom\'etri\-ques du calcul infinit\'esimal, Partie 3, Gauthier-Villars 1894

\noindent[DD1] T. Despeyrous, G. Darboux, Cours de m{\'e}canique, Tome 1, Hermann, Paris (1884)

\noindent[DD2] T. Despeyrous, G. Darboux, Cours de m{\'e}canique, Tome 2, Hermann, Paris (1886)

\noindent[Fau] Faure R., Transformations conformes en m{\'e}canique ondulatoire, {\it Comptes Rendus Acad.\ Sci.\ Paris}, 237 (1953), pp.\ 603--605

\noindent[Fa2] Faure R., Transformation conforme en m{\'e}canique ondulatoire. G{\'e}n{\'e}ralisa\-tion de la
notion de valeur propre. {\it Proceedings of the International Congress of Mathematicians,} vol.\ II (1954), pp.\ 337--339

\noindent[Fu1] Funk P., \"Uber Fl\"achen mit lauter geschlossenen geod\"atischen Linien, {\it Math.\ Ann.}, 74 (1913), pp.\ 278--300

\noindent[Gau] Gauss K.F., Disquisitiones generales circa superficies curvas, G\"ottingen (1828)

\noindent[Gou] Goursat E., Les transformations isogonales en M{\'e}canique, {\it Comptes Rendus Acad.\ Sci.\ Paris}, 108 (1889), pp.\ 446--448

\noindent[Gui] Guicciardini N., The development of Newtonian calculus in Britain 1700--1800, Cambridge University Press (1989)


\noindent[Hal] Halphen G.-H., Sur les lois de Kepler, {\it Bulletin de la Soci\'et\'e Philomatique de Paris}, 7-1 (1878), pp.\ 89--91; {\it \OE uvres 2}, Paris: Gauthier-Villars, 1918, pp.\ 93--95

\noindent[Ha2] Halphen, G.-H., Sur une formule d'analyse. Bull. Soc.\ Math.\ Fr., 8 (1880), pp.\ 62--64

\noindent[Hig] Higgs, P.W., Dynamical symmetries in a spherical geometry, I, J.\ Phys.\ A: Math.\ Gen., 12 (1979), pp.\ 309--323

\noindent[Jov] Jovanovi\'c, V., A note on the proof of Bertrand's theorem, Theoretical and Applied Mechanics 42.1 (2015), pp.\ 27--33

\noindent[Kas] Kasner, E., Differential-geometric aspects of dynamics. The Princeton Colloquium: Lectures on Mathematics, Delivered September 15 to 17, 1909, Before Members of the American Mathematical Society in Connection with the Summer Meeting Held at Princeton University, Volume 3 of American Mathematical Society Colloquium lectures, Part 1, 1913, 1934, Chelsea, 1980

\noindent[Kep] Kepler, J., Epitomes Astronomi\ae\ Copernican\ae, liber quartus (1622)

\noindent[KepE] Kepler, J., Epitome of Copernican astronomy; and, harmonies of the world; translated by Charles Glenn Wallis, Annapolis, St.\ John's Bookstore (1939)

\noindent[Kil] Killing W, Die Mechanik in den Nicht-Euklidischen Raumformen, {\it Journal f{\"u}r die reine und angewandte Mathematik}, 98 (1885), pp.\ 1--48

\noindent[Koe] G. Koenigs, Sur les lois de force centrale fonction de la distance pour laquelle toutes les trajectoires sont alg{\'e}briques, Bulletin de la S.M.F., 17 (1889), pp.\ 153--155

\noindent[KF1] Kudryavtseva, E.A., Fedoseev, D.A., Mechanical systems with closed orbits on manifolds of revolution,
Sbornik: Mathematics 206:4 (2015), pp.\ 718--737


\noindent[KF2] Kudryavtseva, E.A., Fedoseev, D.A., The Bertrand's manifolds with equators, {\it Mos\-cow Univ.\ Math.\ Bull.}, 71 (2016), pp.\ 23--26


\noindent[KF3] Kudryavtseva, E.A., Fedoseev, D.A., Superintegrable Bertrand natural mechanical systems, {\it Journal of Mathematical Sciences}, 248 (2020), pp.\ 409--429

\noindent[Kus] Kustaanheimo, P., Spinor Regularization of the Kepler Motion, Annales Universitatis Turkuensis, A.\ I.\ 73 (1964), 7 pp.

\noindent[KSt] Kustaanheimo, P., Stiefel, E., Perturbation theory of Kepler motion based on spinor regularization, Journal f\"ur die reine und angewandte Mathematik, 218 (1965), pp.\ 204--219

\noindent[Lag] J.-L. Lagrange, Solution de diff\'erens probl{\^e}mes de calcul int{\'e}gral, Miscellanea Taurinensia, Tome 3, (1766), 2$^{\rm e}$ pag., pp.\ 179--380

\noindent[Lar] J. Larmor, On the Immediate Application of the Principle of Least Action to Dynamics of a Particle, Catenaries, and other related Problems, Proceedings of the London Mathematical Society, 15 (1883), pp.\ 158--170

\noindent[Leg] A.M. Legendre, {\it Exercices de calcul int\'egral sur divers ordres de transcendantes et sur les quadratures}, tome second, V$^{\rm e}$ Courcier, Paris (1817), sixi\`eme partie

\noindent[Leg2] A.M. Legendre, {\it Trait\'e des fonctions elliptiques et des 
int\'egrales
eul\'erien\-nes, avec des tables pour en faciliter le calcul 
num\'erique}, tome premier,
Huzard-Courcier, Paris (1825)

\noindent[Lev] T. Levi-Civita, Sulle trasformazioni delle equazioni dinamiche, Annali di Matematica 2-24 (1896), pp.\ 255--300

\noindent[Le1] T. Levi-Civita, Traiettorie singolari ed urti nel problema ristretto dei tre corpi, Annali di Matematica Pura ed Applicata, 3-9 (1904), pp.\ 1--32

\noindent[Le2] T. Levi-Civita, Sur la r\'esolution qualitative du probl\`eme restreint des trois corps, {\it Verhandlungen des dritten Internationalen Mathematiker-Kongresses in Heidelberg vom 8.\ bis 13.\ august 1904}, Leipzig, 1905, pp.\ 402--408

\noindent[Le3] T. Levi-Civita, Sur la r\'esolution qualitative du probl\`eme restreint des trois corps, Acta Mathematica 30.1 (1906), pp.\ 305-327

\noindent[Le4] T. Levi-Civita, Sulla regolarizzazione del problema piano dei tre corpi, Atti della Reale Accademia Dei Lincei, Rendiconti, serie 5, v.\ 24 (1915), pp.\ 61--75

\noindent [Le5] T. Levi-Civita, Sopra due trasformazioni canoniche desunte dal moto parabolico, Atti della Reale Accademia Dei Lincei, Rendiconti, serie 5, v.\ 25 (1916), pp.\ 445--458

\noindent[Le6] T. Levi-Civita, Sur la r\'egularisation du probl\`eme des trois corps, Acta Mathematica, 42 (1920), pp.\ 99--144

\noindent [Le7] T. Levi-Civita, Regolarizzazione del problema dei tre corpi e sua portata, {\it Questioni di Meccanica Classica e Relativistica}, Zanichelli, Bologna (1924), pp.\ 1--38
 
\noindent[Lio] J. Liouville,
Sur un th{\'e}or{\`e}me de M.\ Gauss concernant le produit des deux rayons de courbure principaux en chaque point d'une surface,
Journal de Math{\'e}matiques Pures et Appliqu{\'e}es (1) 12 (1847), pp.\ 291--304

\noindent[Mac] MacLaurin C., {\it Geometria Organica: sive descriptio linearum curvarum universalis}, Gul.\ \& Joh.\ Innys, London, 1720

\noindent[Ma2] MacLaurin C., {\it Treatise of Fluxions. In two books.} Ruddimans, Edinburgh, 1742

\noindent[MaT] Matveev, V.S. and Topalov, P.J., Trajectory Equivalence and
Corresponding Integrals, 
Regular and Chaotic Dynamics, 3 (1998), pp.\ 30--45

\noindent[McG] R. McGehee, Double collisions for a classical particle system with nongravitational interactions, Commentarii Mathematici Helvetici, 56 (1981), pp.\ 524--557

\noindent[Moe] R. Moeckel, Embedding the Kepler Problem as a Surface of Revolution, Regular and Chaotic Dynamics, 23 (2018), pp.\ 695--703

\noindent[Mon] R. Montgomery, Metric cones, N-body collisions, and Marchal's lemma, (2018), arXiv:1804.03059

\noindent [Mos] J.\ Moser, Regularization of Kepler's problem and the averaging method on a manifold, Communications on pure and applied mathematics, 23 (1970), pp.\ 609--636.

\noindent[Nee] T. Needham, Newton and the Transmutation of Force, The American Mathematical Monthly, 100:2 (1993), pp.\ 119--137


\noindent[NeP] A. Nersessian and G. Pogosyan, Relation of the oscillator and Coulomb systems on spheres and pseudospheres, Physical Review A, 63 (2001), 020103(R)

\noindent[New] I. Newton, Philosophi\ae\ Naturalis Principia Mathematica, London, 1687

\noindent[NewE] I. Newton, The {\it Principia}, Mathematical Principles of Natural Philosophy,
A New translation by I.B.\ Cohen and A.\ Whitman,
University of
California Press (1999)

\noindent[Pai] P. Painlev\'e, M\'emoire sur la transformation des \'equations de la Dynamique, J.\ de math\'ematiques pures et appliqu\'ees 10 (1894), pp.\ 5--92

\noindent[Pa1] P. Painlev\'e, Le{\c c}ons sur l'int\'egration des \'equations diff\'erentielles de la m\'ecanique et applications, A.\ Hermann, Paris, 1895

\noindent[Pa2] P. Painlev\'e, Sur les transformations des \'equations de la Dynamique, {\it Comptes Rendus Acad.\ Sci.\ Paris}, 123 (1896), pp. 392--395

\noindent[Per] V. Perlick, Bertrand spacetimes, Class.\ Quantum Grav.\ 9 (1992), pp.\ 1009--1021

\noindent[Pfi] A. Pfister, Die
geod\"atischen Linien einer Klasse
von Fl\"achen, deren Linienelement
den Liouvilleschen Typus hat, Inaugural-Dissertation, G\"ottingen, 1904

\noindent[RLC] G. Ricci, T. Levi-Civita, M\'ethodes de calcul diff\'erentiel absolu et leurs applications, Mathematische Annalen, 54.1 (1900), pp.\ 125--201

\noindent[Rou] E.J. Routh, A Treatise on Dynamics of a Particle: With Numerous Examples, Cambridge University Press, 1898

\noindent[San] M. Santoprete, Gravitational and harmonic oscillator potentials on surfaces of revolution, Journal of Mathematical Physics, 49, 042903 (2008)



\noindent[Ser] Serret P., {\it Th{\'e}orie nouvelle g{\'e}om{\'e}trique et m{\'e}canique des lignes {\`a} double courbure}, Mallet-Bachelier, Paris, 1859

\noindent [Sie] C.L.\ Siegel, Vorlesungen \"uber Himmelsmechanik, Springer-Verlag (1956)

\noindent [SiM] C.L.\ Siegel, J.K.\ Moser, Lectures on Celestial Mechanics, Springer-Verlag (1971)

\noindent[Sun] K.F. Sundman, Recherches sur le probl{\`e}me des trois corps, Acta Societatis Scientiarum Fennic\ae, 34-6 (1907), pp.\ I--II--1--43

\noindent[Su2] K.F. Sundman, Nouvelles recherches sur le probl{\`e}me des trois corps, Acta Societatis Scientiarum Fennic\ae, 35-9 (1909), pp.\ 1--27

\noindent[Su3] K.F. Sundman, M\'emoire sur le probl{\`e}me des trois corps, Acta Mathematica, 36 (1912), pp.\ 105--179

\noindent[Tan] J. Tannery, Sur une surface de r{\'e}volution du quatri{\`e}me degr{\'e} dont les lignes g{\'e}od{\'e}si\-ques sont alg{\'e}briques, Bull.\ Sci.\ Math., Paris (1892), pp.\ 190--192

\noindent[Tho] T.Y. Thomas, {\it On the transformation of the equations of dynamics}, Journal of mathematics and physics 25 (1946),
pp.\ 191--208

\noindent[Zag] O.A. Zagryadskii, Bertrand surfaces with a pseudo-Riemannian metric of revolution, Moscow Univ. Math. Bull., 70 (2015), pp.\ 49--52


\noindent[ZKF] O.A. Zagryadski\v{\i}, E. A. Kudryavtseva and D. A. Fedoseev, A generalization of Bertrand's theorem to surfaces of revolution, Sbornik: Math., 203:8 (2012), pp.\ 1112--1150


\noindent[Zha] L. Zhao, Kustaanheimo-Stiefel Regularization and the Quadrupolar Conjugacy, Regular and Chaotic Dynamics, 20 (2015), pp.\ 19--36

\noindent[Zol] O. Zoll, Ueber Fl\"achen mit Scharen geschlossener geod\"atischer Linien, {\it Math.\ Ann}.\ 57 (1903), pp.\ 108--133

\end{document}